\documentclass[nohyper]{JHEP3}
\usepackage{amsmath}
\usepackage{amsfonts}
\usepackage{slashed}
\usepackage{graphicx}
\usepackage{verbatim}
\usepackage{subfigure}

\newcommand \E {\text{E}}
\newcommand \NE {\text{NE}}

\title{Factorization Properties of Soft Graviton Amplitudes}

\author{Chris D. White\\
School of Physics and Astronomy, Scottish Universities Physics Alliance, University of Glasgow, Glasgow G12 8QQ, 
Scotland, UK\\
 E-mail: \email{c.white@physics.gla.ac.uk}}

\abstract{We apply recently developed path integral resummation 
methods to perturbative quantum gravity. In particular, we provide 
supporting evidence that eikonal graviton amplitudes factorize into hard
and soft parts, and confirm a recent hypothesis that soft gravitons 
are modelled by vacuum expectation values of products of certain 
Wilson line operators, which differ for massless and massive particles. 
We also investigate terms which break this 
factorization, and find that they are subleading with respect to the
eikonal amplitude. The results may help in understanding the connections
between gravity and gauge theories in more detail, as well as 
in studying gravitational radiation beyond the eikonal approximation.}


\begin{document}
\section{Introduction}
\label{introduction}
Perturbative quantum gravity has been well-studied over many decades,
in the form of general relativity minimally coupled to matter particles
(see~\cite{Hamber:2007fk} for a recent review), as well as alternative
theories.
Although the pitfalls of such an approach are well-known -- chiefly that
quantum general relativity contains non-renormalizable ultraviolet
divergences -- it is nevertheless useful for a number of reasons.
Firstly, one may treat perturbative quantum gravity as an effective theory,
and calculate loop corrections to gravitational observables. Secondly, it 
may well be the case that if a consistent quantum theory of gravity is found,
it shares features with quantized general relativity. This motivates the study
of features in this theory which, it may be argued, should be generically 
shared by alternatives. A third reason for studying gravity is that there are 
intriguing connections between the structure of perturbative scattering
amplitudes in gravity and non-abelian
gauge theories (e.g.~\cite{Hohm:2011dz,Bern:2010yg,Bern:2010ue,Bern:2002kj,
Bern:1999ji,ArkaniHamed:2008yf}). A possible 
application of these ideas is, for example, that of settling the question of 
whether $N=8$ supergravity is ultraviolet finite~\cite{Bern:1998ug,
Bern:2006kd,Bern:2007hh,Bern:2008pv,Bern:2009kd}.

In light of the above remarks (and as stressed recently 
in~\cite{Naculich:2011ry}), an interesting property of perturbative
gravity is the structure of infrared singularities. One might hope, for
example, that the infrared behaviour of alternative (possibly UV finite)
quantum gravity theories shares at least some of the features that underly
quantum GR coupled to matter. Furthermore, the structure of IR singularities 
in renormalizable gauge theories is well-understood. IR divergences 
cancel between real and virtual contributions for sufficiently inclusive
observables~\cite{Bloch:1937pw}. The singularity structure nevertheless
becomes relevant in that it affects residual large
contributions to the perturbative expansion of physical cross-sections which
must be resummed to all orders in the coupling constant in order to obtain
sensible physical predictions. Resummation is by now a highly developed 
subject in both abelian~\cite{Yennie:1961ad} 
and non-abelian gauge theories. A number of approaches exist, such as 
Feynman diagram methods~\cite{Sterman:1986aj,Catani:1989ne}, 
Wilson line techniques~\cite{Korchemsky:1992xv,Korchemsky:1993uz}, 
effective field theories~\cite{Beneke:2002ph,Bauer:2000yr,Bauer:2002nz,
Becher:2006nr,Becher:2006mr,Becher:2007ty},
and the recently developed path integral approach of~\cite{Laenen:2008gt}.
Crucial to all these approaches is the notion of {\it factorization}, 
namely that scattering amplitudes separate into a {\it hard interaction},
which is infrared finite, and a {\it soft function}, 
which contains all infrared
singularities. These are generated by the emission of soft (zero momentum)
gauge bosons between the external lines of the amplitude, referred to in
the literature as the {\it eikonal approximation}\footnote{When amplitudes
involve massless external particles, hard collinear singularities must 
also be taken into account in so-called 
{\it jet functions}~\cite{Mueller:1979ih,Collins:1980ih,Sen:1981sd,
Korchemsky:1988pn,Magnea:1990zb}. These are 
not usually considered in gravity, as collinear 
singularities cancel after summing over all emitting 
particles~\cite{Weinberg:1965nx}.}. Such factorization properties are 
important both for resummation applications, and also for understanding the
structure of infrared singularities to all 
orders~\cite{Dixon:2008gr,Gardi:2009qi,Dixon:2009gx,Becher:2009cu,
Becher:2009qa,Becher:2009kw,Dixon:2009ur,Dixon:2010zz,Gardi:2009zv}.

The same physics also governs the infrared properties of perturbative 
gravity amplitudes, as first analysed in~\cite{Weinberg:1965nx}. Recently,
Naculich and Schnitzer have reconsidered the structure of gravitational 
IR divergences~\cite{Naculich:2011ry}. The main point of their paper is to
argue that infrared singularities at all orders are generated by the 
exponentiation of the one-loop divergence i.e. that there are no subleading
divergences. Their argument is based upon the assumption that gravitational 
amplitudes factorize in the same way that gauge theory amplitudes do, in
terms of hard and soft functions. That is, the $n$-graviton
scattering amplitude (adopting the notation of~\cite{Naculich:2011ry}) 
may be written
\begin{equation}
A_n=S_n\cdot H_n,
\label{Andef}
\end{equation}
where $H_n$ is IR-finite, and $S_n$ collects all IR singularities, generated
by soft graviton exchange. They make the further hypothesis that, by analogy
with gauge theory, the soft function can be expressed as a vacuum expectation
value:
\begin{equation}
S_n=\left\langle0\left|\prod_{i=1}^n\Phi_{p_i}(0,\infty)\right|0
\right\rangle,
\label{Snwilson}
\end{equation}
containing the Wilson line operators\footnote{These operators describe
the emission of soft gravitons from hard emitting particles, and should not 
be confused with the parallel transport operator of general relativity, 
defined in terms of the Christoffel symbol, which is also sometimes referred
to in Wilson line terms.}
\begin{equation}
\Phi_p(a,b)={\cal P}\exp\left(i\kappa\int_a^b ds\,p_{i\mu} p_{i\nu}
h^{\mu\nu}(sp)\right),
\label{wilsondef}
\end{equation}
where there is one such operator associated with each external line of the
amplitude (with momentum $p_i$). Here $h^{\mu\nu}$ is the graviton field 
(we will define this more carefully in what follows), and 
$\kappa=32\pi G$, where $G$ is Newton's constant. The line integral 
in eq.~(\ref{wilsondef}) is over a straight line contour, parametrized by 
$s$. 

\begin{figure}
\begin{center}
\scalebox{1.0}{\includegraphics{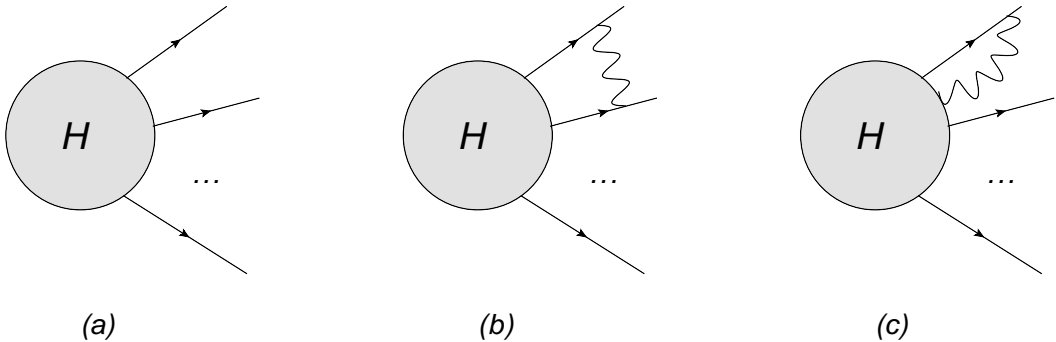}}
\caption{Examples of (a) a generic hard interaction with outgoing particles, 
which may emit soft gauge bosons; (b) an external emission contribution; 
(c) an internal emission contribution.}
\label{intext}
\end{center}
\end{figure}

The aim of this paper is to examine these hypotheses within the path integral
resummation framework of~\cite{Laenen:2008gt}, which was developed in the 
context of abelian and non-abelian gauge theory. The main purpose 
of that paper was to address corrections to the eikonal approximation, and
the result was a systematic classification of the next-to-eikonal contributions
to scattering amplitudes -- that is, those which occur at first 
subleading order in an expansion of the amplitude in the momenta of 
emitted gauge bosons. 
The results were subsequently confirmed using Feynman diagrammatic methods
in~\cite{Laenen:2010uz}, and have also been used to classify the structure 
of (non-abelian) soft gluon corrections in multiparton processes, 
generalising the concept of webs from two parton 
scattering~\cite{Gatheral:1983cz,Frenkel:1984pz,Sterman:1981jc} to the 
multiparton case~\cite{Gardi:2010rn,Gardi:2011wa}\footnote{See 
also~\cite{Mitov:2010rp} for an alternative viewpoint on multiparton webs. 
Other work on next-to-eikonal corrections can be found 
in~\cite{Laenen:2008ux,Moch:2009hr,Soar:2009yh,Vogt:2010pe,Vogt:2010ik,
Grunberg:2009yi,Grunberg:2009vs,Grunberg:2010sw}.}. The results 
of~\cite{Laenen:2008gt} show that the structure of a scattering 
amplitude ${\cal A}$ subject to soft photon emissions has the schematic form
\begin{equation}
{\cal A}={\cal A}_0\exp\left[\sum_{G^{\E}}G^{\E}+\sum_{G^{\NE}}G^{\NE}
\right]\left(1+
{\cal A}_{rem.}\right).
\label{ampstruc}
\end{equation}
Here ${\cal A}_0$ is the hard interaction amplitude undressed by soft 
photons, and the exponent contains connected subdiagrams (which span 
the external lines) $G^{\E}$ and $G^{\NE}$ at eikonal and next-to-eikonal 
order respectively. A set of effective Feynman rules for forming these 
diagrams has been given in~\cite{Laenen:2008gt,Laenen:2010uz}, which 
generalize the well-known eikonal Feynman rules to subleading order in 
the momentum expansion. The term ${\cal A}_{rem.}$ in eq.~(\ref{ampstruc}) 
is also next-to-eikonal order, but does not formally 
exponentiate. It consists of diagrams in which eikonal photons are 
emitted from an external line and land inside the hard interaction. We 
may call these {\it internal emission} diagrams to distinguish them 
from the {\it external emission} diagrams which enter the exponent, and 
their origin makes clear that they can be thought of directly as a breaking 
of the factorization of the amplitude into hard and soft parts. Although 
internal emission contributions do not formally exponentiate, they do have 
an iterative structure to all orders in perturbation theory, which is 
fixed by gauge invariance by a result known as the {\it Low-Burnett-Kroll} 
theorem~\cite{Low:1958sn,Burnett:1967km} (see also~\cite{DelDuca:1990gz} 
for a further generalization). Examples of internal and
external emission contributions are shown in figure~\ref{intext}.

Although the case of abelian gauge theory is described in the preceding 
paragraph, a very similar structure occurs in non-Abelian theories 
(essentially, the sum over connected diagrams in eq.~(\ref{ampstruc}) 
is replaced by a sum over so-called {\it webs}, as described 
in~\cite{Gatheral:1983cz,Frenkel:1984pz,Sterman:1981jc,
Mitov:2010rp,Gardi:2010rn,Gardi:2011wa}). 
We will see in this paper that the same general 
structure is also observed in perturbative gravity, assuming that the 
infrared properties can be analysed independently of the uncertain UV 
completion of the theory. In particular, we will derive directly the 
appropriate form of the Wilson line operator which describes soft
graviton emissions, as well as the appropriate gravitational generalization of 
Low's theorem, which shows that internal emission contributions in gravity 
are subleading with respect to the eikonal approximation, as is also the 
case in gauge theory. These results strengthen the hypotheses made 
in~\cite{Naculich:2011ry}, and may also shed more light on
the correspondence between gauge theory and gravity. 

There are also phenomenological motivations for studying corrections to
eikonal gravity. The latter has been used in a variety of recent applications,
in particular the study of transplanckian scattering in different
gravitational theories. This has a number of potential uses, such as 
investigating whether different gravity theories have the same long distance
behaviour~\cite{Giddings:2010pp}, or even what potential collider signatures
are in extra dimension scenarios~\cite{Stirling:2011mf}. The classification
of corrections to the eikonal approximation, as examined in this paper,
may allow further study of these and related subjects.

The structure of the paper is as follows. In section~\ref{sec:path} we 
describe the path integral framework for scalar particles emitting soft 
gravitons, adapting the approach used for abelian gauge theory
in~\cite{Laenen:2008gt}, and derive the form of the Wilson line operator 
for soft gravitons. In section~\ref{sec:low} we discuss how gauge 
invariance can be used to constrain factorization-breaking terms, and 
discuss how this relates to Low's theorem~\cite{Low:1958sn} in QED. 
In section~\ref{sec:discuss} we discuss our results before concluding.

\section{Path integral approach to soft graviton amplitudes}
\label{sec:path}
In this section, we will consider the scattering amplitude for $L$ scalar 
particles, dressed by any number of soft gravitons. Although pure 
multigraviton amplitudes were considered in~\cite{Naculich:2011ry}, 
scalar particles will be sufficient for our purposes, 
due to the fact that eikonal gravitons are insensitive to the spin 
of the particle which emits them~\cite{Weinberg:1965nx}.

The action for general relativity minimally coupled to a (complex) scalar field
is given by
\begin{equation}
S_{\rm grav}=S_{\rm E.H.}[g^{\mu\nu}]+S_{\rm mat}[\phi^*,\phi,g^{\mu\nu}],
\label{Stot}
\end{equation}
here $S_{\rm E.H.}$ is the Einstein-Hilbert action (which must be suitably 
gauge-fixed in order to define the graviton propagator), and\footnote{Note 
that we use the metric (-,+,+,+), as in~\cite{Laenen:2008gt}.}
\begin{equation}
S_{\rm mat}[\phi^*,\phi,g^{\mu\nu}]=\int d^d x\sqrt{-g}\left[-g^{\mu\nu}
\partial_\mu\phi^*\partial_\nu\phi-m^2\phi^*\phi\right],
\label{Smatdef}
\end{equation}
in $d$ dimensions, where $g$ is the determinant of the metric tensor 
$g^{\mu\nu}$. Perturbation theory can then be defined after expanding the 
metric tensor about the flat space Minkowski metric $\eta^{\mu\nu}$. As 
explained in e.g.~\cite{Hamber:2007fk}, there is an ambiguity in how one 
performs the weak field expansion. Here we adopt the approach 
of~\cite{Capper:1973bk}, and define
\begin{equation}
\tilde{g}^{\mu\nu}=\sqrt{-g}g^{\mu\nu}.
\label{gtildedef}
\end{equation}
This introduces a Jacobian in principle in eq.~(\ref{Smatdef}), although 
this may be taken to be one in dimensional 
regularization~\cite{Hamber:2007fk}\footnote{Many other field
redefinitions are possible instead of that of eq.~(\ref{gtildedef}).
See~\cite{Hohm:2011dz} for a recent proposal in the context of exploring
the factorization of gravity amplitudes into gauge theory amplitudes.}. 
We then define the graviton field $h^{\mu\nu}$ via
\begin{equation}
\tilde{g}^{\mu\nu}=\eta^{\mu\nu}+\kappa h^{\mu\nu},
\label{hdef}
\end{equation}
where $\kappa^2=32\pi G$ as in~\cite{Naculich:2011ry}. Note that symmetry
of the metric tensor implies $h^{\mu\nu}=h^{\nu\mu}$. One could also have 
chosen, of course, to expand the metric $g^{\mu\nu}$ directly. However, 
the choice of eq.~(\ref{hdef}) results in simpler expressions for the 
scalar-graviton vertices. We will return to this
point later on when discussing next-to-eikonal corrections. 

With the above definition for $h^{\mu\nu}$ one has~\cite{Capper:1973bk}
\begin{equation}
\sqrt{-g}=1+\frac{\kappa}{d-2}h^{\alpha}_{\alpha}+\kappa^2\left[\frac{(
h^{\alpha}_{\alpha})^2}{(d-2)^2}-\frac{h^{\alpha\beta}h_{\alpha\beta}}
{d-2}\right]+{\cal O}(\kappa^3).
\label{gexpand}
\end{equation}
Furthermore, the inverse of $\tilde{g}^{\mu\nu}$ is given 
by~\cite{Capper:1973pv}
\begin{equation}
\tilde{g}_{\mu\nu}=\eta_{\mu\nu}-\kappa h_{\mu\nu}+\kappa^2
h_\mu^{\phantom{\nu}\alpha}h_{\alpha\nu}+{\cal O}(\kappa^2).
\label{gexpand2}
\end{equation}
The action of eq.~(\ref{Smatdef}), evaluated to quadratic order in $\kappa$, 
is thus
\begin{align}
S_{\rm mat}[\phi^*,\phi,h^{\mu\nu}]&=\int d^d x\left\{-\eta^{\mu\nu}
\partial_\mu\phi^*\partial_\nu\phi-\kappa h^{\mu\nu}\partial_\mu\phi^*
\partial_\nu\phi\right.\notag\\
&\left.\quad-m^2\phi^*\phi-\frac{m^2\kappa}{d-2}\left[h^\alpha_\alpha+
\kappa\left(\frac{(h^\alpha_\alpha)^2}{(d-2)}-h^{\alpha\beta}
h_{\beta\alpha}\right)\right]\phi^*\phi\right\}.
\label{Smatdef2}
\end{align}
For what follows it is useful to write this as
\begin{equation}
S=-\int d^dx\phi^*\hat{S}\phi,
\label{Shat}
\end{equation}
where the quadratic operator $\hat{S}$ is given by
\begin{align}
\hat{S}&=-\eta^{\mu\nu}\partial_\mu\partial_\nu-\kappa(\partial_\mu h^{\mu\nu})
\partial_\nu-\kappa h^{\mu\nu}\partial_\mu\partial_\nu+m^2\notag\\
&\quad+\frac{m^2\kappa}
{d-2}\left[h^\alpha_\alpha+\kappa\left(\frac{
(h^\alpha_\alpha)^2}{(d-2)}-h^{\alpha\beta}h_{\beta\alpha}\right)\right]
\label{Shat2}
\end{align}
(n.b. we have integrated by parts where necessary, neglecting surface terms). 

The starting point of the path integral approach of~\cite{Laenen:2008gt} is to 
separate the gauge field (here the graviton) into hard and soft modes. That is,
in the path integral which defines the quantum gravity theory, one may write
\begin{equation}
\int{\cal D}h^{\mu\nu}=\int{\cal D}h^{\mu\nu}_h{\cal D}h^{\mu\nu}_s,
\label{hsep}
\end{equation} 
where $h^{\mu\nu}_h$ and $h^{\mu\nu}_s$ contain only hard and soft modes in 
momentum space respectively. The exact details of this definition (which 
may proceed e.g. by constructing an explicit hypersurface in the 
multigraviton momentum space which separates hard and soft parts) need not 
concern us, as discussed in~\cite{Laenen:2008gt} for the QED case. However, 
we note that this separation breaks the full gauge invariance
of the gravity theory. Under a gauge transformation, the graviton behaves as
\begin{equation}
h^{\mu\nu}(x)\rightarrow h^{\mu\nu}(x)+\partial^\mu\xi^\nu(x)
+\partial^\nu\xi^\mu(x)
\label{htrans}
\end{equation}
where $x$ denotes 4-position. A general gauge function $\xi^\mu$ 
(defined in position space) will in general have both soft and hard modes 
in momentum space, so that gauge transformations exist which violate the 
separation into soft and hard modes introduced in eq.~(\ref{hsep}).
However, a residual gauge invariance remains, namely that one may 
transform $h^{\mu\nu}$ in momentum space according to any 
transformation\footnote{We use the same symbol to denote
the position- and momentum-space graviton fields, where the argument 
of the field removes any ambiguity.}
\begin{equation}
h_{h,s}^{\mu\nu}(k)\rightarrow h_{h,s}^{\mu\nu}(k)+k^\mu\xi_{h,s}^\nu(k)
+k^\nu\xi_{h,s}^\mu(k)
\label{htrans2}
\end{equation}
in which the gauge function $\xi^\mu_{h,s}$ contains only hard or soft modes 
as required. This symmetry will be sufficient to derive the structure of 
factorization-breaking corrections from internal emission graphs in 
section~\ref{sec:low}.

We now consider the Green's function for the scattering of $L$ scalar 
particles which, after performing the above separation of the graviton 
field, may be written
\begin{equation}
G(p_1\ldots,p_L)=\int{\cal D}h^{\mu\nu}_s H(x_1,\ldots, x_L)
e^{iS_{\rm E.H.}[h^{\mu\nu}_s]}\prod_{j=1}^L\langle p_j|(\hat{S}
-i\epsilon)^{-1}|x_j\rangle.
\label{Greens}
\end{equation}
Here $H(x_1,\ldots,x_L)$ is the {\it hard interaction}, which produces scalar 
particles at 4-positions $\{x_i\}$, whose external momenta are 
given by $\{p_i\}$. This is directly analagous to the hard interaction given 
for the QED case in~\cite{Laenen:2008gt}, except for the fact that this 
will now consist of a sum of Feynman diagrams involving hard graviton 
modes, rather than hard photon modes. There are also integrations
over the positions $\{x_i\}$, which we do not show explicitly in 
eq.~(\ref{Greens}). Associated with each external line 
is a propagator for a scalar particle in a background soft gravitational 
field. According to the well-known definition of the propagator, this
is given by the inverse of the quadratic operator $\hat{S}$ for the scalar 
field $\phi$, which includes the relevant gravitational interactions as 
shown in eq.~(\ref{Shat2}). The propagator is sandwiched between states 
of given initial position and final momentum. The latter is due to the fact 
that Green's functions (and, consequently, scattering amplitudes) are 
usually considered in momentum space. The former arises because, in what 
follows, we will interpret soft graviton scattering in 
terms of the space-time worldlines of the emitting scalars which participate
in the hard interaction.

The next step is to note that the propagator factors in eq.~(\ref{Greens})
can be expressed in terms of first quantized path integrals. Such a technique
arose some years ago~\cite{Strassler:1992zr,Schmidt:1994zj,vanHolten:1995ds}, 
motivated by string theoretic approaches to field theory 
amplitudes~\cite{Bern:1991an,Bern:1991aq} (see also~\cite{Karanikas:2002sy}
for an example of worldline techniques applied in a resummation context).
Here we shall quote the result as written in~\cite{Laenen:2008gt}, that the 
propagator factors appearing in eq.~(\ref{Greens}) may be written
\begin{align}
\langle p_j|(\hat{S}-i\epsilon)^{-1}|x_j\rangle&=\frac{1}{2}\int_0^\infty dT
\int_{x(0)=x_j}^{p(T)=p_j}{\cal D}p{\cal D}x\exp\left[\phantom{\int}
-ip(T)\cdot x(T)\right.\notag\\
&\left.\quad+i\int_0^Tdt(p\cdot\dot{x}-\hat{H}(p,x))\right],
\label{proppath}
\end{align}
where
\begin{equation}
\hat{H}=\frac{1}{2}\hat{S}.
\label{Hdef}
\end{equation}
Some explanatory comments are in order. Firstly, there is a double path 
integral over the position and momentum trajectories of the particle, 
which are parametrized in terms of a time-like variable $t$ (the upper 
limit of $T$ will eventually be taken
to $\infty$, corresponding to the final state). The boundary conditions 
are that the particle must be produced at position $x_j$ ($x(0)=x_j$), and 
end up with final momentum $p_j$ ($p(T)=p_j$). Equation~(\ref{proppath}) 
arises as the solution of a Schr\"{o}dinger equation for a certain 
evolution operator formed out of $\hat{S}$, and indeed has the 
recognizable Feynman path integral representation, where the
second term in the exponent is the classical action formed out of 
the ``Hamiltonian'' $\hat{H}$. The first term in the exponent arises due 
to the fact that we are sandwiching the propagator between position and 
momentum states, rather than two position states. Note that the path 
integral over $x$ in eq.~(\ref{proppath}) has a clear physical 
interpretation in terms of a sum over all possible spacetime 
trajectories of the scalar particle whose final momentum is $p_j$. 
This will be crucial to isolating the properties of the eikonal 
approximation (and beyond) in what follows.

From eqs.~(\ref{Shat2}) and~(\ref{Hdef}), we find that the appropriate 
Hamiltonian operator for a scalar particle coupled to gravity is
(dropping the subscript $s$ on the soft graviton field $h^{\mu\nu}_s$, which
we do from now on)
\begin{align}
\hat{H}&=-\frac{1}{2}\eta^{\mu\nu}\partial_\mu\partial_\nu-\frac{\kappa}{2}
(\partial_\mu h^{\mu\nu})\partial_\nu-\frac{\kappa}{2}h^{\mu\nu}
\partial_\mu\partial_\nu\notag\\
&\quad+\frac{1}{2}m^2+\frac{m^2\kappa}{2(d-2)}\left[h^\alpha_\alpha
+\kappa\left(\frac{
(h^\alpha_\alpha)^2}{(d-2)}-h^{\alpha\beta}h_{\beta\alpha}\right)\right],
\label{Hgrav}
\end{align}
which in momentum space ($p_\mu=i\partial_\mu$)\footnote{Care must be taken
with the sign of the momentum operator, which is here chosen to ensure 
consistency with the diagrammatic calculation of appendix~\ref{app:diags}.} 
becomes
\begin{align}
\hat{H}&=\frac{1}{2}(p^2+m^2)+\frac{\kappa}{2}p_\mu p_\nu h^{\mu\nu}
+\frac{i\kappa}{2}
p_\nu(\partial_\mu h^{\mu\nu})\notag\\
&\quad+\frac{m^2\kappa}{2(d-2)}\left[h^\alpha_\alpha+
\kappa\left(\frac{
(h^\alpha_\alpha)^2}{(d-2)}-h^{\alpha\beta}h_{\beta\alpha}\right)\right],
\label{Hgrav2}
\end{align}
so that the propagator function of eq.~(\ref{proppath}) becomes
\begin{align}
\langle p_j|(\hat{S}-i\epsilon)^{-1}|x_j\rangle&=\frac{1}{2}\int_0^\infty dT
\int_{x(0)=x_j}^{p(T)=p_j}{\cal D}p{\cal D}x\exp\left\{
-ip(T)\cdot x(T)+i\int_0^Tdt\left[\phantom{\int}p\cdot\dot{x}\right.\right.\notag\\
&\quad-\frac{1}{2}(p^2+m^2)-\frac{\kappa}{2}p_\mu p_\nu h^{\mu\nu}
-\frac{i\kappa}{2}p_\nu(\partial_\mu h^{\mu\nu})\notag\\
&\quad\left.\left.-\frac{m^2\kappa}{2(d-2)}\left[h^\alpha_\alpha
+\kappa\left(\frac{
(h^\alpha_\alpha)^2}{(d-2)}-h^{\alpha\beta}h_{\beta\alpha}\right)\right]\right]
\right\}.
\label{proppath2}
\end{align}

At this point, we can recover the eikonal approximation. Recalling 
that this corresponds to the momentum $k$ of any emitted gravitons being 
completely soft ($k\rightarrow0$), this means that the emitting scalar 
particles do not recoil. Thus, they follow the straight line classical 
trajectories
\begin{equation}
x(t)=x_j+p_j t.
\label{classtraj}
\end{equation}
Subeikonal corrections correspond to a systematic expansion about this 
trajectory. That is, for each external line one writes
\begin{equation}
x(t)\rightarrow x_j+p_j t+x(t),\quad p(t)\rightarrow p_j+p(t),
\label{eikexpand}
\end{equation}
where the boundary conditions for the transformed variables are $x(0)=p(T)=0$. 
One then finds~\cite{Laenen:2008gt}
\begin{align}
\langle p_j|(\hat{S}-i\epsilon)^{-1}|x_j\rangle&=\frac{1}{2}\int_0^\infty 
dTe^{-ip_j\cdot x_j-\frac{1}{2}(p_j^2+m^2)T}f(T),
\label{proppath3}
\end{align}
where
\begin{align}
f(T)&=
\int_{x(0)=0}^{p(T)=0}{\cal D}p{\cal D}x\exp\left\{
i\int_0^Tdt\left[p\cdot\dot{x}-\frac{1}{2}p^2-\frac{\kappa}{2}
(p_{j\mu}+p_\mu) (p_{j\nu}+p_\nu) 
h^{\mu\nu}\right.\right.\notag\\
&\left.\left.\quad-\frac{i\kappa}{2}(p_{j\nu}+p_\nu)
(\partial_\mu h^{\mu\nu})
-\frac{m^2\kappa}{2(d-2)}\left[h^\alpha_\alpha
+\kappa\left(\frac{
(h^\alpha_\alpha)^2}{(d-2)}-h^{\alpha\beta}h_{\beta\alpha}\right)\right]\right]
\right\}.
\label{proppath4}
\end{align}
The path integral over $p$ is Gaussian, and may be performed explicitly. First,
one writes eq.~(\ref{proppath4}) as
\begin{equation}
f(T)=\int_{x(0)=0}^{p(T)=0}{\cal D}p{\cal D}x\exp\left[
i\int_0^Tdt\left(-\frac{1}{2}p_\mu A^{\mu\nu}p_\nu+B^\mu p_\mu +C\right)
\right],
\label{proppath5}
\end{equation}
where we have defined
\begin{align}
A^{\mu\nu}&=\eta^{\mu\nu}+\kappa h^{\mu\nu};\notag\\
B^\mu&=\dot{x}^\mu-\kappa p_{j\nu} h^{\mu\nu}
-\frac{i\kappa}{2}\partial_\nu h^{\nu\mu};\notag\\
C&=-\frac{\kappa}{2}p_{j\mu}p_{j\nu}h^{\mu\nu}-\frac{i\kappa}{2}
p_{j\nu}\partial_\mu h^{\mu\nu}-\frac{m^2\kappa}{2(d-2)}
\left[h^\alpha_\alpha+\kappa\left(\frac{
(h^\alpha_\alpha)^2}{(d-2)}-h^{\alpha\beta}h_{\beta\alpha}\right)\right],
\label{ABC}
\end{align}
and also used the fact that $h^{\mu\nu}=h^{\nu\mu}$. 
Using standard results for Gaussian integrals gives
\begin{equation}
f(T)=\int{\cal D}x\exp\left[
i\int_0^Tdt\left(\frac{1}{2}B^\mu (A^{-1})_{\mu\nu}B^\nu +C\right)\right],
\label{fT}
\end{equation}
where we have ignored an overall normalization constant (which is cancelled 
upon correctly normalising the measure for the path integral). 
Noting that $A^{\mu\nu}$ as defined in eq.~(\ref{ABC}) is just the metric 
tensor
$\tilde{g}^{\mu\nu}$ of eq.~(\ref{gtildedef}), its inverse is simply 
$(A^{-1})_{\mu\nu}=\tilde{g}_{\mu\nu}$, as given in eq.~(\ref{gexpand2}). Then
eq.~(\ref{fT}) gives (after some tedious algebra)
\begin{align}
f(T)&=\int{\cal D}x\exp\left\{i\int_0^Tdt\left[\frac{\dot{x}^2}{2}
-\frac{\kappa}{2}\dot{x}^\mu h_{\mu\nu}\dot{x}^\nu-\kappa
p_{j\alpha}h^{\mu\alpha}\dot{x}_\mu
-\frac{i\kappa}{2}(\partial_\alpha h^{\alpha\nu})\dot{x}_\nu\right.
\right.\notag\\
&\left.\left.\quad+\frac{\kappa^2}{2}\dot{x}^\mu\dot{x}^\nu 
h_{\mu}^{\phantom{\mu}\alpha}h_{\alpha\nu}+\kappa^2p_{j\alpha}
h^{\mu\alpha}h_{\mu\nu}\dot{x}^\nu+\frac{i\kappa^2}{2}
(\partial_\beta h^{\beta\nu})h_{\mu\nu}\dot{x}^\mu
+\frac{\kappa^2}{2}p_{j\alpha}p_{j\beta}
h^{\mu\alpha}h_\mu^{\phantom{\mu}\beta}\right.\right.\notag\\
&\left.\left.\quad+\frac{i\kappa^2}{2}p_{j\alpha}
h_\nu^{\phantom{\nu}\alpha}(\partial_\beta h^{\beta\nu})
-\frac{\kappa^2}{8}(\partial_\alpha h^{\alpha\mu})(\partial_\beta 
h^{\beta}_{\phantom{\beta}\mu})-\frac{\kappa}{2}p_{j\mu}p_{j\nu}h^{\mu\nu}
-\frac{i\kappa}{2}p_{j\nu}\partial_\mu
h^{\mu\nu}\right.\right.\notag\\
&\left.\left.\quad-\frac{m^2}{2}\frac{\kappa}{(d-2)}h^\mu_\mu
-\frac{m^2\kappa^2}{2}\left(\frac{(h^\alpha_\alpha)^2}{(d-2)^2}
-\frac{h^{\alpha\beta}h_{\beta\alpha}}{d-2}\right)\right]\right\}.
\label{fT2}
\end{align}

One may relate all this back to the eikonal scattering amplitude as follows.
After combining eq.~(\ref{fT2}) into eq.~(\ref{proppath3}) and 
substituting the resulting propagator factors into the Green's function 
of eq.~(\ref{Greens}), one finds (after truncating the external lines with 
free scalar propagators as required by the LSZ formula) that 
the scattering amplitude for $L$ scalar particles has the form
(see the derivation in~\cite{Laenen:2008gt})
\begin{equation}
{\cal A}(p_1,\ldots,p_L)=\int{\cal D}h^{\mu\nu} H(x_1,\ldots,x_L)
e^{iS_{\rm E.H.}[h^{\mu\nu}]}\prod_{j=1}^{\infty}e^{-ip_j\cdot x_j}
f_j(\infty).
\label{amp1}
\end{equation}
That is, each external line is associated with a factor $f_j(\infty)$, given 
by eq.~(\ref{fT2}) with $T\rightarrow\infty$, which still implicitly 
includes a path integral over all possible trajectories of particle $j$. 
We will shortly analyse this result further, but first note that 
eq.~(\ref{amp1}) has the form of a generating functional for a
quantum field theory, i.e. for the soft graviton field. The external 
line factors act as source terms in this theory which, as can be seen 
from eq.~(\ref{fT2}), generate soft graviton emission vertices 
localized along the external lines. The Feynman diagrams generated
by eq.~(\ref{amp1}) are thus {\it subdiagrams} in the full theory, which 
span the external lines. Given that connected diagrams exponentiate in 
quantum field theory, we can immediately conclude that a large class of soft
graviton corrections (namely those consisting of connected subdiagrams 
formed from the Feynman rules derivable from eq.~(\ref{fT2})) exponentiate. 
This is a generalization of the original result in~\cite{Weinberg:1965nx}, 
which is based solely on the eikonal contributions. The argument given here 
for the exponentiation of soft graviton corrections based on the textbook 
exponentiation properties of connected diagrams in quantum field theory
is essentially exactly the same as the argument used for QED 
in~\cite{Laenen:2008gt}\footnote{Strictly speaking, exponentiation of 
connected diagrams holds only subject to
the condition that the source terms for the gauge field are mutually
commuting. This is true 
for gravity and abelian gauge theories, but is not the case in 
non-Abelian gauge theories, where a more sophisticated treatment of 
exponentiation is then required~\cite{Laenen:2008gt,Gardi:2010rn,
Mitov:2010rp}.}.

Let us now physically interpret the various terms in the external line 
factor of eq.~(\ref{fT2}) in more detail. First, note that the Feynman 
rules generated by each term cannot be immediately read off, due to the 
fact that for each external line the path integral over trajectories 
$x(t)$ has yet to be carried out. However, as shown in~\cite{Laenen:2008gt},
this can be done by systematically expanding about the classical trajectory.
The leading term (as already remarked above) gives the eikonal approximation,
in which the particles do not recoil. The first subleading corrections give
the next-to-eikonal corrections, and further corrections are 
next-to-next-to-eikonal etc. Put another way, an expansion about the classical
trajectories of the hard emitting particles (in position space) is entirely 
equivalent to an expansion in the momenta of the emitted gravitons 
(a momentum space expansion). We may evaluate the eikonal contribution in
eq.~(\ref{fT2}) by setting $x(t)=p(t)=0$, and also neglecting terms which
are quadratic in $h^{\mu\nu}$ (these lead to two-graviton emission vertices,
which are suppressed with respect to the eikonal limit, due to having
one less eikonal denominator compared with two separate graviton emissions).
One may also neglect any terms containing derivatives of the graviton field,
as these (in momentum space) will be suppressed by powers of the graviton
momentum. The result is
\begin{equation}
f^{\rm Eik.}_j(\infty)=\int {\cal D}x\exp\left[i\int_0^\infty dt 
\left(-\frac{\kappa}{2}
p_{j\mu}p_{j\nu}h^{\mu\nu}-\frac{m^2}{2(d-2)}\eta_{\mu\nu}h^{\mu\nu}\right)\right].
\label{fT3}
\end{equation}
Carrying out the path integral over $x$ gives unity assuming the appropriate
normalization, and thus the eikonal approximation for the scattering amplitude
is
\begin{align}
{\cal A}(p_1,\ldots,p_L)&=\int{\cal D}h^{\mu\nu} H(x_1,\ldots,x_L)
e^{iS_{\rm E.H.}[h^{\mu\nu}]}\prod_{j=1}^{\infty}e^{-ip_j\cdot x_j}
\exp\left[i\int_0^\infty dt \left(-\frac{\kappa}{2}
p_{j\mu}p_{j\nu}h^{\mu\nu}\right.\right.\notag\\
&\left.\left.\quad-\frac{m^2}{2(d-2)}\eta_{\mu\nu}h^{\mu\nu}\right)\right].
\label{amp2}
\end{align}
In fact, up to next-to-eikonal corrections, one may set the initial 
positions of the external lines to $x_j=0$~\cite{Laenen:2008gt}, so that
eq.~(\ref{amp2}) simplifies further to
\begin{align}
{\cal A}(p_1,\ldots,p_L)&=H(0,\ldots,0)\int{\cal D}h^{\mu\nu} 
e^{iS_{\rm E.H.}[h^{\mu\nu}]}\prod_{j=1}^{\infty}
\exp\left[i\int_0^\infty dt \left(-\frac{\kappa}{2}
p_{j\mu}p_{j\nu}h^{\mu\nu}\right.\right.\\
&\left.\left.\quad-\frac{m^2}{2(d-2)}\eta_{\mu\nu}h^{\mu\nu}\right)\right],
\label{amp3}
\end{align}
where we have taken the hard interaction outside the path integral over the 
soft gauge field. One thus sees that the amplitude has factorized explicitly
into a hard and a soft part, where the soft gravitons are described by 
the Wilson line factors
\begin{equation}
\Phi_j=\exp\left[i\int_0^\infty dt \left(-\frac{\kappa}{2}
p_{j\mu}p_{j\nu}h^{\mu\nu}-\frac{m^2}{2(d-2)}\eta_{\mu\nu}h^{\mu\nu}
\right)\right].
\label{Wilson1}
\end{equation}
The generating functional for the soft graviton field theory indeed generates
vacuum expectation values of these operators, as expressed 
in eq.~(\ref{Snwilson}). Let us now discuss the physics
of these operators in more detail.

Firstly, the first term in the exponent in eq.~(\ref{Wilson1}) agrees,
up to a normalization factor, with the hypothesized Wilson
line operator of~\cite{Naculich:2011ry} (which dealt with pure graviton
scattering, and thus massless external particles). 
Note, however, that the Feynman rule
for the soft graviton generated by this term is derived from
\begin{equation}
-\frac{i\kappa}{2}\int_0^\infty dt p_{j\mu}p_{j\nu}h^{\mu\nu}(p_jt)=
\int\frac{d^dk}{(2\pi)^d}\left(\frac{\kappa}{2}\frac{p_\mu p_\nu}
{p\cdot k}h^{\mu\nu}(k)\right),
\label{feynrule}
\end{equation}
where $h^{\mu\nu}(k)$ is the momentum space graviton field. The
momentum space eikonal Feynman rule is thus
\begin{displaymath}
\frac{\kappa}{2}\frac{p_\mu p_\nu}{p\cdot k},
\end{displaymath}
whose normalization agrees with the result quoted 
in~\cite{Naculich:2011ry}. Thus, 
we maintain that eq.~(\ref{Wilson1}) is correctly normalized.

Secondly, note that there is an additional contribution in the case in which 
the external particles are massive, as has been considered here, which 
appears as the second term in the exponent of eq.~(\ref{Wilson1}). Wilson
line operators in conventional gauge theories are the same for massless and
massive particles, due to the fact that the mass does not source the 
gauge field. Here, however, the mass plays the role of a gravitational 
charge, and thus we expect such a dependence. Furthermore, such a 
term can be motivated on general Lorentz invariance grounds, given that
the exponent of the Wilson line operator can only depend on the properties
of the emitting particle. That is, the prefactor which one contracts with
the graviton field can only depend upon $p^\mu$ and $m$, and the only two
combinations having two Lorentz indices and the correct mass dimension are
\begin{displaymath}
p_\mu p_\nu,\quad m^2\eta_{\mu\nu},
\end{displaymath}
which indeed both occur.

Thirdly, it is interesting to note the behaviour of eq.~(\ref{Wilson1})
under rescalings of external line momentum. In QED, the Wilson line operator
has the form
\begin{equation}
\Phi_{\rm QED}=\exp\left[ie\int dt p_\mu A^\mu\right],
\label{Wilson2}
\end{equation}
where $e$ is the electromagnetic charge. This leads to the momentum-space
eikonal Feynman rule
\begin{equation}
\frac{p_\mu}{p\cdot k}
\label{feynQED}
\end{equation}
for emission of a soft photon of momentum $k$, and this latter expression
is manifestly invariant under the transformation $p\rightarrow\lambda p$,
corresponding to the invariance of eq.~(\ref{Wilson2}) under reparametrizations of
the Wilson line. By contrast, the eikonal Feynman rules generated from
eq.(\ref{Wilson1}) do not have this property. Rather, they can be written
in the form
\begin{equation}
\frac{\kappa}{2}\left[p_{\mu}\left(\frac{p_\nu}{p\cdot k}\right)
+m\left(\frac{\eta_{\mu\nu}}{2(d-2)}\frac{m}{p\cdot k}\right)\right].
\label{feyngrav2}
\end{equation}
The quantities in the round brackets are both manifestly scale invariant
under rescalings of the external momentum. These are then multiplied 
respectively by factors of $p_\mu$ and $m$, namely by gravitational charges.
Thus, the eikonal Feynman rules in gravity have the form of a scale-invariant
(in $p$) quantity multiplied by a charge, which is exactly what one expects
by analogy with abelian and non-abelian gauge theories. 

The curious reader may wonder about the presence of poles in $d-2$ 
in the Wilson line operators. As discussed in e.g.~\cite{Hamber:2007fk}, 
perturbation theory becomes singular in $d=2$ dimensions, owing to the fact 
that the Einstein-Hilbert action is then a topological invariant (i.e. there 
is no non-trivial dynamics). Whether or not such poles occur in the Wilson 
line operators is dependent on the choice of weak field approximation. If
one expands the metric itself rather than absorbing $\sqrt{-g}$, the $(d-2)$
poles instead occur in the graviton propagator, such that one can never
get rid of them, as expected from their physical origin.

Above, we have confirmed the factorization property of soft graviton 
amplitudes in the eikonal approximation, in terms of vacuum expectation
values of appropriate Wilson line operators. Some comments are in order
regarding the plethora of additional terms in eq.~(\ref{fT2}). As
explained in~\cite{Laenen:2008gt}, the path integral over $x$ in
eq.~(\ref{fT2}) can be systematically expanded about the leading
eikonal term. This was done for the case of abelian and non-abelian
gauge theory by defining $p=\lambda n$, where $\lambda\rightarrow\infty$
isolates the eikonal term, and $n^2=0$ or $n^2=1$ for massless and massive 
particles respectively.
The path integral could then be done perturbatively
in $\lambda^{-1}$ (also taking care to expand the argument of each gauge boson
field), leading to a set of next-to-eikonal Feynman rules.
There were two types of next-to-eikonal vertex: those involving a single
photon or gluon emission, and those involving two gauge bosons emitted from
the same position on the external line. These vertices were then confirmed
by an explicit diagrammatic treatment in~\cite{Laenen:2010uz}. 

A similar procedure is possible in the present case. 
Note, however, that the simple 
scaling $p_j=\lambda n_j$ does not solely isolate the eikonal term as
$\lambda\rightarrow\infty$. Instead, eq.~(\ref{fT2}) becomes 
(as $\lambda\rightarrow\infty$)
\begin{align}
\lim_{\lambda\rightarrow\infty}f(T)&=\int{\cal D}x\exp\left\{i\int_0^Tdt\left[
-\frac{\kappa\lambda^2}{2}n_{j\mu}n_{j\nu}h^{\mu\nu}
-\frac{\lambda^2m^2\kappa}{d-2}\eta_{\mu\nu}h^{\mu\nu}
+\frac{\lambda^2\kappa^2}{2}n_{j\alpha}n_{j\beta}
h^{\mu\alpha}h_\mu^{\phantom{\mu}\beta}\right.\right.\notag\\
&\left.\left.\quad
-\frac{\lambda^2m^2\kappa^2}{2}\left(\frac{(h^\alpha_\alpha)^2}{(d-2)^2}
-\frac{h^{\alpha\beta}h_{\beta\alpha}}{d-2}\right)\right]\right\}.
\label{fTeik}
\end{align}
The first two terms are the eikonal contribution of eq.~(\ref{Wilson1}),
whereas the last two term are two-graviton vertices, which have no 
analogue in the case of abelian or non-abelian gauge theory. They are clearly
suppressed in the eikonal limit, as such vertices have one less eikonal
denominator than two separate eikonal emission vertices. The above remarks
imply that the $\lambda$ scaling of~\cite{Laenen:2008gt} 
is not quite correct as a systematic method for expanding about the eikonal
approximation\footnote{We will see in what follows, however, that
the third term in eq.~(\ref{fTeik}) is cancelled upon performing
the path integral over the trajectory $x$.}, 
although does indeed lead to correct results for the gauge 
theories studied in that paper. The extra terms arise here essentially because
of the weak field expansion, which allows the possibility to construct 
terms which are quadratic in external momenta or masses, 
and quadratic (or higher) in the graviton field. Such terms are absent in
QED, where no weak field expansion is needed. 
In any case, it is straightforward to 
isolate next-to-eikonal contributions from power-counting, 
even if one does not use the $\lambda$-expansion (which 
amounts to little more than a convenient 
book-keeping device where applicable). To calculate the next-to-eikonal
graviton Feynman rules, one must perform the path integral perturbatively
as explained in appendix B of~\cite{Laenen:2008gt}. Here we perform this
calculation in appendix~\ref{app:path}, with the result that, up to
next-to-eikonal order, the external line factor of eq.~(\ref{fT2}) may be 
written (taking $T\rightarrow\infty$)
\begin{align}
f(\infty)&=\exp\left\{\int\frac{d^dk}{(2\pi)^d}h^{\mu\nu}(k)
\left[\frac{\kappa}{2}\frac{p_\mu p_\nu}{p\cdot k}-\frac{\kappa}{4}
\frac{p_{(\mu}k_{\nu)}}{p\cdot k}+\frac{m^2}{2}\frac{\kappa}{(d-2)}
\frac{\eta_{\mu\nu}}{p\cdot k}+\frac{\kappa}{4}\frac{k^2}{(p\cdot k)^2}
p_\mu p_\nu\right.\right.\notag\\
&\left.\left.+\frac{m^2}{4}\frac{\kappa}{d-2}\frac{k^2}{(p\cdot k)^2}
\eta_{\mu\nu}\right]+\int\frac{d^d k}{(2\pi)^d}\int\frac{d^dl}{(2\pi)^d}
h^{\mu\nu}(k)\,h^{\alpha\beta}(l)\left[-\frac{\kappa^2}{16}
\left(\frac{p_\alpha \,p_\beta \,p_{(\mu}k_{\nu)}}{p\cdot k \,p\cdot (k+l)}
\right.\right.\right.\notag\\
&\left.\left.\left.
+\frac{p_\alpha \,p_\beta \,p_{(\mu}\,l_{\nu)}}{p\cdot l\,p\cdot (k+l)}
+(\mu\nu\leftrightarrow\alpha\beta)\right)+\frac{\kappa^2}{8}
\frac{p_\mu\,p_\nu\,p_\alpha\,p_\beta\,(k\cdot l)}{p\cdot k\,p\cdot l\,
p\cdot(k+l)}-\frac{m^2\kappa^2}{16(d-2)}\left(\frac{\eta_{\mu\nu}\,
p_{(\alpha}\,k_{\beta)}}{p\cdot k\,p\cdot(k+l)}\right.\right.\right.\notag\\
&\left.\left.\left.+\frac{\eta_{\mu\nu}\,
p_{(\alpha}\,l_{\beta)}}{p\cdot l\,p\cdot(k+l)}+(\mu\nu\leftrightarrow\alpha
\beta)\right)+\frac{m^2\kappa^2}{8(d-2)}\left(\frac{\eta_{\mu\nu}\,p_\alpha\,
p_\beta\,(k\cdot l)}{p\cdot k\,p\cdot l\,p\cdot(k+l)}+(\mu\nu
\leftrightarrow\alpha\beta)\right)\right.\right.\notag\\
&\left.\left.+\frac{m^4\kappa^2}{8(d-2)^2}\frac{\eta_{\mu\nu}\,\eta_{\alpha
\beta}\,(k\cdot l)}{p\cdot k\,p\cdot l\,p\cdot(k+l)}+\frac{m^2\kappa^2}{4
p\cdot(k+l)}\left(\frac{2\eta_{\mu\nu}\,\eta_{\alpha\beta}}{(d-2)^2}
-\frac{1}{d-2}(\eta_{\mu\beta}\,\eta_{\nu\alpha}+\eta_{\nu\beta}\,
\eta_{\mu\alpha})\right)\right]\right\}.
\label{expmom}
\end{align}
Here we have written the exponent explicitly in momentum space, where
we use the argument of the graviton field to identify its 
Fourier-transform.
From eq.~(\ref{expmom}) one can easily read off the effective 
Feynman rules for graviton emission up to NE order. There are eikonal and 
NE one-graviton vertices given by
\begin{equation}
\scalebox{0.6}{\includegraphics{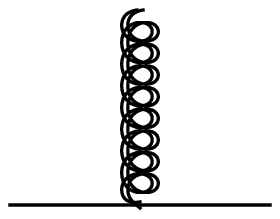}}\quad\frac{\kappa}{2}\frac{1}
{p\cdot k}\left[p_\mu p_\nu+\frac{m^2}{d-2}\eta_{\mu\nu}\right]
\label{1gErule}
\end{equation}
and
\begin{equation}
\scalebox{0.6}{\includegraphics{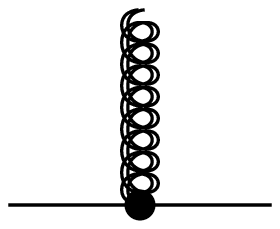}} 
\quad\frac{\kappa}{4}\left[
-\frac{p_{(\mu}k_{\nu)}}{p\cdot k}+\frac{k^2}{(p\cdot k)^2}
\left(p_\mu \,p_\nu+\frac{m^2}
{d-2}\eta_{\mu\nu}\right)\right]
\label{1gNErule}
\end{equation}
respectively, where the eikonal result for massless particles
is already well-known~\cite{Weinberg:1965nx}. There is also an 
effective two-graviton vertex, given by
\begin{align}
\scalebox{0.6}{\includegraphics{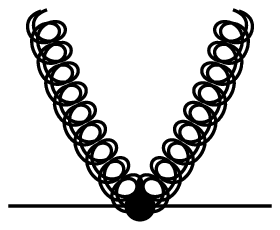}}
\quad&\frac{\kappa^2}{16}\frac{1}{p\cdot k\,p\cdot l\,p\cdot(k+l)}\left\{
-\left[\left(p_\alpha p_\beta+\frac{m^2}{d-2}\eta_{\alpha\beta}\right)
\left(p_{(\mu}k_{\nu)}(p\cdot l)+p_{(\mu}l_{\nu)}(p\cdot k)\right)\right.
\right.\notag\\
&\left.\left.\quad
-\frac{2m^2}{d-2}\eta_{\mu\nu}p_\alpha p_\beta(k\cdot l)+(\mu\nu
\leftrightarrow\alpha\beta)\right]+2p_\mu\,p_\nu\,p_\alpha\,p_\beta(k\cdot l)
+\frac{2m^4}{(d-2)^2}\eta_{\mu\nu}\eta_{\alpha\beta}(k\cdot l)\right.\notag\\
&\left.\quad
+4m^2(p\cdot k)(p\cdot l)\left(\frac{2\eta_{\mu\nu}\eta_{\alpha\beta}}
{(d-2)^2}-\frac{1}{d-2}\left(\eta_{\mu\beta}\eta_{\nu\alpha}+\eta_{\nu\beta}
\eta_{\mu\alpha}\right)\right)\right\}.
\label{2gNErule}
\end{align}
One expects such a vertex for two reasons. Firstly, whilst successive graviton
emissions are completely decoupled in the eikonal approximation, this is not 
expected to hold beyond the eikonal approximation, such that one expects
two-graviton correlations. Secondly, there is an exact two-graviton vertex in
the theory (as discussed in appendix~\ref{app:diags}), which forms part of the
result of eq.~(\ref{2gNErule}). It is useful to cross-check 
the above results using diagrammatic methods,
as has been carried out for the case of QED and QCD in appendix B 
of~\cite{Laenen:2008gt}. We present such a calculation for the present case
in appendix~\ref{app:diags} of this paper.

The above results imply that a subset of next-to-eikonal corrections 
exponentiates at NE order in perturbative quantum gravity. Namely,
connected external emission graphs formed from effective eikonal
and NE Feynman rules, where each graph contains at most one NE vertex.
Some further comments are in order regarding the results we have obtained.

To start with, one may question how useful the NE Feynman rules are in 
practice. Eikonal Feynman rules tell us about infrared singularities, 
which may be expected to be somewhat universal in alternative field 
theories of gravity. Next-to-eikonal corrections are non-singular, and thus 
potentially tell us less about generic infrared features of gravity theories.
 Furthermore, the fact that perturbative GR coupled to matter is not UV 
renormalizable suggests that an expansion in the momenta of emitted gravitons
may cease to be meaningful at some point. Thus, one should be wary of 
taking any NE corrections too seriously. What matters in the above analysis
is merely that in the soft limit, we have shown that the eikonal approximation
leads to Wilson line operators as hypothesized in~\cite{Naculich:2011ry}, 
where factorizable corrections are genuinely 
subleading in that they do not produce
IR singularities. We have also clarified the situation when external particles
have non-zero mass.
 
Another important point is that unlike in abelian and non-abelian gauge 
theory, the NE Feynman rules
are not unique. They depend on the nature of the weak field approximation
used to define the graviton field (e.g. whether one expands $g^{\mu\nu}$
or $\tilde{g}^{\mu\nu}$). This is not the case for the eikonal terms 
in four dimensions. We have verified by explicit calculation (which we do
not consider worth recording here) that the 
Wilson line operator which occurs upon expanding $g^{\mu\nu}$ rather than
$\tilde{g}^{\mu\nu}$ is
\begin{equation}
\Phi_j=\exp\left[i\int_0^\infty dt \left(-\frac{\kappa}{2}
p_{j\mu}p_{j\nu}h^{\mu\nu}-\frac{m^2}{4}\eta_{\mu\nu}h^{\mu\nu}
\right)\right].
\label{Wilson3}
\end{equation}
Comparing this with eq.~(\ref{Wilson2}), we see that the two Wilson line
operators are equal for $d=4$. They are also automatically equal for massless
external particles. The next-to-eikonal terms, however, would differ between
the two calculations, due to extra terms originating from the expansion of
$\sqrt{-g}$ in the kinetic energy term for the scalar field.
Such differences presumably cancel at the level of
full diagrams (as the choice of weak field approximation shuffles 
contributions between different vertices), but it is still true that a level
of ambiguity occurs in thinking about NE Feynman rules that is absent in
conventional gauge theories, as a weak field expansion is not necessary in the
latter.   

A further comment is in order regarding the weak field approximation.
In the above analysis, we have performed two expansions. Firstly,
we have separated the graviton field into hard and soft modes, and used this
as a basis for an expansion in the momenta of emitted gravitons. Secondly, 
we have expanded the graviton field itself in the gravitational coupling
constant $\kappa$. In the latter expansion, we kept terms only up to
${\cal O}(\kappa^2)$. How can we be sure that by including higher order terms,
we do not induce further (next-to) eikonal corrections? 
To see that this is not the 
case, it suffices to note that any power of $\kappa$ is accompanied by a 
further power of the graviton field $h^{\mu\nu}$, whose Lorentz indices
must be appropriately contracted with other factors of $h^{\mu\nu}$,
$\dot{x}^\mu$, $p_j^\mu$, $p^\mu$ or $\partial^\mu$ in eq.~(\ref{fT2}).
This always results in a multigraviton vertex, which is suppressed with
respect to the appropriate number of eikonal 1-graviton vertices due to the 
fact that it lacks at least one eikonal denominator. The fact that 
all vertices at ${\cal O}(\kappa^n)$ generate $n$-graviton vertices also
shows that for next-to-eikonal corrections it is sufficient to expand only
up to ${\cal O}(\kappa^2)$, as we have done above. This is itself interesting,
as it shows that the momentum expansion and weak field expansions are 
partially correlated in a well-defined sense (although there is, of course, 
a tower of subleading momentum terms at any given order in $\kappa$). 

\begin{figure}
\begin{center}
\scalebox{1.0}{\includegraphics{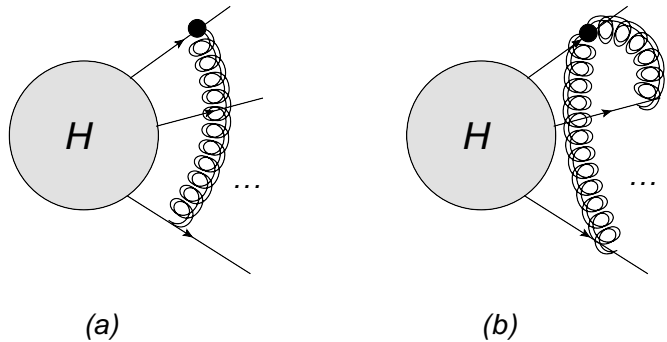}}
\caption{Example external emission graphs which exponentiate at 
next-to-eikonal order, where $\bullet$ represents a next-to-eikonal
vertex, the doubled gluon line a graviton, and all other vertices are
assumed to be eikonal.}
\label{NEexp}
\end{center}
\end{figure}

It follows from the above discussion that the NE Feynman rules will
contain 1-graviton and 2-graviton vertices.  It then follows (as for
the eikonal corrections) that a large class of next-to-eikonal graviton
corrections exponentiates, namely those graphs containing connected 
subdiagrams involving one next-to-eikonal Feynman rule. Examples are
shown in figure~\ref{NEexp}.
Thus, the exponent of the scattering amplitude for $L$-particle scattering
dressed by soft gravitons has the same schematic form as the QED case of
eq.~(\ref{ampstruc}). As in that case, this is not the whole story. One
may also get additional contributions from internal emission diagrams such
as that shown in figure~\ref{intext}(c), which explicitly break the
factorization of the scattering amplitude into hard and soft parts. These
are known to be subleading in the soft limit in conventional gauge 
theories~\cite{Low:1958sn,Burnett:1967km,DelDuca:1990gz}. It is likely,
but not perhaps obvious, that the same is true for gravity. This is the subject
of the following section.

\section{Factorization breaking terms}
\label{sec:low}
In the previous section, we applied the path integral resummation method 
of~\cite{Laenen:2008gt} to perturbative general relativity minimally coupled
to a complex scalar field. We found that the amplitude factorizes into hard
and soft parts, where the latter is decribed by a vacuum expectation value of
Wilson line operators as in eq.~(\ref{Wilson1}). Explicit calculation 
yields diagrams formed out of eikonal Feynman rules, which then exponentiate. 
Corrections to the exponent are subleading, and consist of diagrams containing
one next-to-eikonal Feynman rule, with all other vertices eikonal. The aim of 
this section is to examine what happens when this factorization is broken,
and to argue that, as in QED, such corrections are next-to-eikonal, thus do 
not affect the structure of IR singularities. 

First let us briefly recap what happens in QED. We denote by 
$H^{\mu}(p_1,\ldots p_L;k)$ the subamplitude for production of a soft photon
emerging from the hard interaction $H(p_1,\ldots p_L)$ with momentum $k$ and 
with Lorentz index $\mu$. Then gauge invariance may be applied to show 
that $H^\mu$ is related to the hard interaction itself, up to next-to-eikonal 
order. More specifically, this relation has the following form:
\begin{equation}
H^\mu(p_1,\ldots p_L;k)=-\sum_{j=1}^Lq_j\frac{\partial}{\partial
p_{j\mu}} H(p_1,\ldots p_L),
\label{low1}
\end{equation} 
where $q_j$ is the electric charge of particle $j$. This equation tells us
that factorization breaking terms are subleading with respect to the
eikonal approximation. Furthermore, their structure is completely fixed
up to next-to-eikonal order. Such contributions do not formally exponentiate
(as do external emission graphs), but they do nevertheless have an iterative
structure to all orders in perturbation theory, as given by eq.~(\ref{low1}).
This equation is an important ingredient in Low's 
theorem~\cite{Low:1958sn,Burnett:1967km,DelDuca:1990gz}, which relates the
vertex function in gauge theory to the vertex function with an additional
emission of a gauge boson. 

The above result suggests that we should attempt a similar strategy in quantum 
gravity. That is, one may impose gauge invariance on the factorized form
for the soft graviton scattering amplitude, and check that this leads to 
extra contributions which are suppressed with respect to the eikonal limit.
As discussed in the previous section, the separation of the graviton
field into hard and soft modes has broken the full gauge invariance of the
theory. Nevertheless, the residual gauge symmetry under transformations 
of the type shown in eq.~(\ref{htrans2}) will be sufficient to derive a 
gravitational analogue of eq.~(\ref{low1}).

Our starting point is the factorized form of the soft graviton amplitude
of eq.~(\ref{amp2}), where each external line factor is reduced to a
Wilson line operator. Under a gauge transformation (eq.~(\ref{htrans})), 
the Wilson operators transform as
\begin{equation}
f_j(h^{\mu\nu}+\partial^\mu\xi^\nu+\partial^\nu\xi^\mu)
\rightarrow f(h^{\mu\nu})e^{-i\kappa p_j\cdot\xi_j}.
\label{ftrans}
\end{equation}
Requiring that eq.~(\ref{amp2}) be gauge invariant then implies that
the hard interaction transforms as
\begin{equation}
H(x_1,\ldots x_L;h^{\mu\nu}+\partial^\mu\xi^\nu+\partial^\nu\xi^\mu)
\exp\left[i\kappa\sum_jp_j\cdot\xi(x_j)\right].
\label{Htrans}
\end{equation}
Expanding both sides to first order in $h^{\mu\nu}$ and $\xi^\mu$ gives
\begin{align}
&\quad H(x_1,\ldots,x_L)+\int d^dxH_{\mu\nu}(x_1,\ldots x_L;x)\left[
h^{\mu\nu}(x)+\partial^\mu\xi^\nu(x)+\partial^\nu\xi^\mu(x)\right]\notag\\
&=H(x_1,\ldots x_L)+\int d^dxH_{\mu\nu}(x_1,\ldots,x_L;x)h^{\mu\nu}(x)\notag\\
&\quad+i\kappa\int d^dx\left[H(x_1,\ldots x_L)\sum_j p_{j\mu}
\delta(x-x_j)\right]\xi^\mu(x),
\label{Hexp1}
\end{align}
where all hard functions are evaluated with the untransformed graviton 
field $h^{\mu\nu}$, and $H_{\mu\nu}(x_1,\ldots,x_L;x)$ is the
subamplitude for emission of a graviton from the hard interaction at position
$x$ (this is the analogue of $H^\mu(x_1,\ldots,x_L;x)$ in the QED case, whose
Fourier transform appears on the left-hand side of eq.~(\ref{low1})).
Equation~(\ref{Hexp1}) simplifies to
\begin{align}
&\quad\int d^dxH_{\mu\nu}(x_1,\ldots x_L;x)\left[
\partial^\mu\xi^\nu(x)+\partial^\nu\xi^\mu(x)\right]
\notag\\
&=i\kappa\int d^dxH(x_1,\ldots x_L)\sum_j p_{j\mu}\delta(x-x_j)\xi^\mu(x),
\label{Hexp2}
\end{align}
and integrating by parts on the left-hand side gives
\begin{equation}
-\int d^dx\xi^\mu\partial^\nu\bar{H}_{\mu\nu}
=i\kappa\int d^dxH(x_1,\ldots x_L)\sum_j p_{j\mu}\delta(x-x_j)\xi^\mu(x),
\label{Hexp3}
\end{equation}
where $\bar{H}_{\mu\nu}=H_{\mu\nu}+H_{\nu\mu}$. Due to the fact that
$\xi^\nu$ is arbitrary, one may write
\begin{equation}
-\partial^\nu\bar{H}_{\mu\nu}=i\kappa\sum_jp_{j\mu}H(x_1,\ldots x_L)
\delta(x-x_j).
\label{Hexp4}
\end{equation}
In momentum space this becomes
\begin{equation}
-ik^\nu\bar{H}_{\mu\nu}=i\kappa\sum_jp_{j\mu}H(p_1,\ldots,p_j+k,\ldots p_L),
\label{Hexpmom1}
\end{equation}
and expanding this to ${\cal O}(k)$ gives
\begin{equation}
-k^\nu\bar{H}_{\mu\nu}=\kappa\sum_j\left[p_{j\mu}+ p_{j\mu}k^\nu
\frac{\partial}{\partial p_j^\nu}\right]H(p_1,\ldots,p_L).
\label{Hexpmom2}
\end{equation}
The zeroth order term on the right-hand side cancels due to momentum
conservation
\begin{equation}
\sum_jp_{j\mu}=0,
\label{momcon}
\end{equation}
and one finally finds
\begin{equation}
\bar{H}_{\mu\nu}=-\kappa\sum_jp_{j\mu}\frac{\partial}{\partial p_j^\nu}
H(p_1,\ldots,p_L).
\label{gravlow}
\end{equation}
This is the gravitational equivalent of the QED result expressed by 
eq.~(\ref{low1}). It forms part of a gravitational version of Low's theorem,
and describes the result of internal emission diagrams analagous 
to that shown in 
figure~\ref{intext}(c), in which a soft graviton emerges from the hard 
interaction and lands on an external line. Clearly such diagrams are
next-to-eikonal from eq.~(\ref{gravlow}). To see this, imagine starting
with an external emission graph, and taking one of the soft gravitons off 
an external line and into the hard interaction. Such a replacement replaces
an eikonal Feynman rule with the contribution of eq.~(\ref{gravlow}), and
the resulting diagram thus has one less eikonal denominator of form 
$p\cdot k$, therefore is subleading with respect to the eikonal approximation.
It is worth noting that one may also derive eq.~(\ref{gravlow}) from a 
traditional diagrammatic treatment - we present an example calculation
in appendix~\ref{app:low}.

There is also an additional contribution at NE order, corresponding to 
the fact that the external lines do not start at the origin, but have 
non-zero initial 4-positions $x_j$, as taken into account by the factors
$e^{-p_j\cdot x_j}$ in eq.~(\ref{amp1}). As in~\cite{Laenen:2008gt}, one may
evaluate this contribution by writing the eikonal one-graviton source
term for an external line of momentum $p$ as
\begin{equation}
\frac{\kappa}{2}\int\frac{d^dk}{(2\pi)^d}\frac{h^{\mu\nu}}{p\cdot k}\left(
p_\mu\,p_\nu+\frac{m^2}{d-2}\eta_{\mu\nu}\right)e^{ix\cdot k}
=\frac{\kappa}{2}\int\frac{d^dk}{(2\pi)^d}\frac{h^{\mu\nu}}{p\cdot k}\left(
p_\mu\,p_\nu+\frac{m^2}{d-2}\eta_{\mu\nu}\right)(1+ix\cdot k),
\label{jterm1}
\end{equation}
where we have expanded to NE order on the right-hand side. Combining the
second term on the right-hand side with the rest of eq.~(\ref{gravlow}),
the scattering amplitude is given by
\begin{align}
&{\cal A}(p_1,\ldots,p_L)=\int{\cal D}h\,e^{iS_{\rm E.H.}[h]}\left[
\left(-\sum_{j=1}^L p_{j\mu}\frac{\partial}{\partial p_j^\nu}H(p_1\ldots,p_L)
\right)\int\frac{d^dk}{(2\pi)^d}h^{\mu\nu}\right.\notag\\
&\left.+\int dx_1^d\ldots dx_L^d
H(x_1,\ldots x_n)\left(\sum_j\int\frac{d^dk}{(2\pi)^d}
(ix_j\cdot k)\frac{h^{\mu\nu}}{p_j\cdot k}\left(p_{j\mu}\,p_{j\nu}
+\frac{m^2}{d-2}
\eta_{\mu\nu}\right)\right)e^{-i\sum_jx_j\cdot p_j}
\right]\notag\\
&\times\prod_{j=1}^N f(0,p_j;h),
\label{jterm2}
\end{align}
where we have explicitly reinstated the integrals over the initial positions
of the external lines, and $f(0,p_j;h)$ is the external line factor starting
at the origin for line $j$. Performing the position integrals, one obtains
\begin{align}
{\cal A}(p_1,\ldots,p_L)&=\int{\cal D}h\left\{\int\frac{d^dk}{(2\pi)^d}
h^{\mu\nu}(k)\sum_{j=1}^L\left[-\left(p_{j\mu}\,p_{j\nu}+\frac{m^2}{d-2}
\eta_{\mu\nu}\right)k^\sigma\frac{\partial}{\partial p_j^\sigma}
-p_j^\mu\frac{\partial}{\partial p_j^\nu}\right]\right.\notag\\
&\left.\quad\times H(p_1,\ldots,p_L)
\right\}\prod_{j=1}^N f(0,p_j;h).
\label{jterm3}
\end{align}
One sees that the result of the non-zero initial positions of the
external lines is to add an extra NE term
involving momentum derivatives of the hard interaction. A similar result occurs
in QED and QCD, and it is useful to physically interpret this extra 
contribution. In abelian and non-abelian gauge theory, massless particles
give rise to collinear as well as soft singularities, and one may write the
scattering amplitude in the factorized form (analagous to eq.~(\ref{Andef}))
\begin{equation}
{\cal A}=H\cdot S\cdot\prod_{j=1}^L J_j.
\label{ampfac}
\end{equation}
Here $H$ and $S$ are the hard and soft functions, where the latter collects
all singularities due to the emitted gauge boson momenta satisfying 
$k\rightarrow 0$. Furthermore, $J_j$ is a {\it jet function}, and contains
all singularities arising from hard emissions which are collinear to 
the outgoing particle of momentum $p_j$\footnote{Note 
there is an overlap between
$S$ and $J_j$, namely when singularities become soft and collinear. This is 
usually dealt with by dividing either the soft or jet functions by the 
{\it eikonal jet functions} evaluated in the soft limit, which removes any
double counting.}. Del Duca has shown~\cite{DelDuca:1990gz} that 
in such cases there are extra contributions to the Low-Burnett-Kroll theorem 
due to emissions of soft gluons from inside the jet functions. This is
indeed the physical meaning of the additional term in eq.~(\ref{jterm3}).
One would not ordinarily write jet functions in perturbative gravity, due to
the fact that hard collinear singularities cancel after summing over all
external particle lines and using momentum conservation~\cite{Weinberg:1965nx}.
However, if one is interested in the structure of next-to-eikonal corrections,
there are such terms which do indeed come from emissions within jet functions, 
and which do not cancel after summing over parton legs (due presumably to the
fact that the additional term in eq.~(\ref{jterm3}) is quadratic in each 
external momentum). It may thus be useful to consider jet functions in gravity
after all.

In this section, we have investigated the structure of terms which break the
factorization of soft graviton amplitudes into hard and soft parts, and found
contributions whose physical meaning is that of emission of soft gravitons from
inside the hard and jet functions. All such contributions are subleading with
respect to the eikonal approximation, thus verifying the factorization of 
graviton ampitudes at eikonal order. Furthermore, the structure of soft
graviton amplitudes up to NE order has the same schematic form as in QED or
QCD, namely that of eq.~(\ref{ampstruc}). External emission graphs (consisting
of connected subdiagrams formed out of eikonal and next-to-eikonal 
Feynman rules) formally exponentiate, whereas internal emission graphs have
an iterative structure and do not formally enter the exponent.

\section{Discussion}
\label{sec:discuss}
In this paper, we have studied the recent hypothesis made 
in~\cite{Naculich:2011ry}, that soft graviton amplitudes factorize into
hard and soft functions, as do gauge theory amplitudes. Our analysis
relied on path integral resummation techniques developed in the context
of abelian and non-abelian gauge theory~\cite{Laenen:2008gt}, which allow
an efficient classification of the structure of scattering amplitudes
including subleading corrections with respect to the eikonal approximation.
We find that gravity amplitudes have the same schematic structure as in gauge
theory, as written here in eq.~(\ref{ampstruc}). That is, eikonal graphs
exponentiate (as already known~\cite{Weinberg:1965nx}), and next-to-eikonal
graphs can be separated into external and internal emission contributions.
External emission graphs are connected subdiagrams formed out of effective
eikonal and next-to-eikonal Feynman rules, whose form we derived in this paper.
Internal emission graphs consist of soft gravitons which are emitted from
inside the hard interaction, or jet functions associated with each external 
line. External emission graphs formally exponentiate, whereas internal emission
graphs have an iterative structure to all orders in perturbation theory. 
Importantly, internal emission graphs are subleading with respect to the 
eikonal approximation. 

Note that, as in the cases of QED and QCD considered in~\cite{Laenen:2008gt},
one may quibble whether it is really valid to separate internal and external
emission contributions as far as exponentiation is concerned, or even what
is meant by next-to-eikonal exponentiation at all. After all, upon performing
the exponent of the NE external emission contributions, any term involving
products of two or more NE terms is NNE order or higher, such that 
eq.~(\ref{ampstruc}) is formally equivalent to
\begin{equation}
{\cal A}={\cal A}_0\exp\left[\sum_{G^{\E}}G^{\E}
\right]\left(1+\sum_{G^{\NE}}G^{\NE}
+{\cal A}_{rem.}\right),
\label{ampstruc2}
\end{equation}
in which formula external and internal emission contributions are on the
same footing. Nevertheless, it is true that the former genuinely exponentiate,
whereas the latter have an iterative structure and thus must be calculated 
differently. Furthermore, either eq.~(\ref{ampstruc}) or eq.~(\ref{ampstruc2})
allow NE contributions to be calculated to all orders in perturbation theory,
even if this is really a consequence of the exponentiation of eikonal terms,
as shown explicitly in eq.~(\ref{ampstruc2}).

Aside from the structure of NE effects, an important question 
regarding these results is: have we formally proved 
the factorization of graviton amplitudes (albeit those involving only
scalar external lines) into hard and soft parts? After all, the path 
integral technique we have used starts off by assuming that we may separate
the gauge field into hard and soft modes, and that the resulting exponential
structure captures all singularities. It may perhaps be prudent to regard
the results we have obtained as being supporting evidence for the factorization
property of amplitudes, rather than a rigorous proof. 
Nevertheless, it is {\it very strong} evidence, for a number
of reasons. Firstly, we have analysed corrections which explicitly break the
factorization property (i.e. internal emission graphs), and found these to be
subleading in the momenta of the emitted gluons. We showed this using both
the path integral framework, and also using diagrammatic methods (although this
is essentially the same calculation, as it is gauge invariance that ultimately 
fixes these contributions). Secondly, we find that the structure of eikonal
and next-to-eikonal contributions is exactly analagous to that of QED and QCD.
This suggests that the qualitative structures one finds in the infrared
limits of both gauge and gravity theories are somewhat generic. 
In gauge theories, the factorization property (together with the structure of
the NE Feynman rules) is indeed explicitly corroborated using diagram-based
methods, which gives us confidence that the same would be true for gravity.
A more formal proof of the factorization of gravity amplitudes may perhaps
proceed using a power-counting analysis based upon the Landau equations
associated with the singularities of Feynman diagrams, as
has been carried out in gauge theories (see e.g.~\cite{Sterman:1995fz}
for a pedagogical review).

We are also able to confirm the hypothesis of~\cite{Naculich:2011ry} that
the soft graviton part of the scattering amplitude is described by a vacuum
expectation value of certain Wilson line operators, with one such operator
associated with each external line. This in itself is not so surprising: in
the eikonal approximation external particles do not recoil, and thus can only
be changed by a phase. If this phase is to have the correct gauge 
transformation properties to slot into an amplitude, then some sort of Wilson
line operator is inevitable. In gravity, the Wilson line operator can only 
depend upon $p_\mu\,p_\nu$ and $m^2\eta_{\mu\nu}$ due to Lorentz invariance.
It is therefore reassuring to have verified that this is indeed the case.
The fact that massive external particles give an additional contribution to
the Wilson line operator is also expected, given that $p_\mu$ and $m$ play
the role of charges in gravity. The exponent should consist of charges
multiplying factors which are invariant under rescalings of the external
momenta, which is indeed the case.

In this paper we have considered scalar particles emitting soft gravitons.
One could also consider the more phenomenologically relevant examples of
fermions, gauge bosons or hard gravitons emitting gravitational radiation.
This should make no difference to the eikonal approximation, which is known
to be insensitive to the spin of the emitting particles~\cite{Weinberg:1965nx}.
Beyond the eikonal approximation, and by analogy with the QED case examined
in~\cite{Laenen:2008gt,Laenen:2010uz} one expects NE corrections which are 
insensitive to the spin of the emitting particle, together with additional
contributions which couple graviton emissions to the spin of the emitter.
For fermions, this would need an appropriate curved space formulation of
the second order fermion action~\cite{Morgan:1995te}, which was used
in~\cite{Laenen:2008gt} to analyse fermions in the path integral approach
to soft gauge boson emission. 

The results presented here provide further insight into the relationships
between gauge theories and perturbative quantum gravity. They may also be
useful in phenomenological applications of gravitational physics e.g. 
extending results obtained in the eikonal approximation for 
transplanckian scattering~\cite{Stirling:2011mf,Giddings:2010pp}. 
The investigation of such higher order effects in both gravity and gauge 
theories is ongoing.

\acknowledgments

CDW thanks Jack Laiho and David Miller for
many useful conversations, and is also grateful to Einan Gardi and Lorenzo
Magnea for discussions, and comments on the manuscript. 
He is supported by the STFC Postdoctoral Fellowship
``Collider Physics at the LHC''. We have used JaxoDraw~\cite{Binosi:2008ig,Binosi:2003yf} 
throughout the paper.

\appendix
\section{NE Feynman rules from the path integral approach}
\label{app:path}
The aim of this appendix is to explicitly carry out the path integral
of eq.~(\ref{fT2}), perturbatively about the classical trajectory.
We will keep terms up to and including next-to-eikonal corrections,
for which a weak field expansion to ${\cal O}(\kappa^2)$ is sufficient.
The calculation closely follows that of the QED case in appendix B 
of~\cite{Laenen:2008gt}.

We begin by noting that the first term in the exponent of eq.~(\ref{fT2})
defines the quadratic operator (after taking $T\rightarrow\infty$)
\begin{equation}
i\int_0^\infty dt\frac{\dot{x}^2}{2}\equiv\int_0^\infty dt x_\mu
\left(-\frac{i}{2}\eta^{\mu\nu}\frac{d^2}{dt^2}\right)x_\nu,
\label{xquad}
\end{equation}
where we have integrated by parts on the right-hand side. The propagator
for the $x_\mu$ coordinate is given by the inverse of this operator, which
is~\cite{Laenen:2008gt}
\begin{equation}
G(t,t')=i\min(t,t')\eta_{\mu\nu}.
\label{xprop}
\end{equation}
This in turn leads to the following correlators, which we will need in what
follows:
\begin{align}
&\langle x_\mu(t)x_\nu(t')\rangle=G(t,t')=i\min(t,t')\eta_{\mu\nu}\notag\\
&\langle \dot{x}_\mu(t)x_\nu(t')\rangle=\frac{\partial G(t,t')}{\partial t}
=i\Theta(t'-t)\eta_{\mu\nu}\label{xcorrels}\\
&\langle\dot{x}_\mu(t)\dot{x}_\nu(t')\rangle=
\frac{\partial^2G(t,t')}{\partial t
\partial t'}=i\delta(t-t')\eta_{\mu\nu},\notag
\end{align}
where $\Theta(t-t')$ is the Heaviside function. Note that the second and third 
correlators vanish at equal times ($t=t'$). To see this, one may use finite
difference formulae consistent with the space-time discretization assumed in
the path integral derivation of~\cite{Laenen:2008gt} to write
\begin{equation}
\langle x_\mu(t)\dot{x}_\nu(t)\rangle=\lim_{\epsilon\rightarrow 0^+}
\frac{\langle x_\mu(t)[x_\nu(t+\epsilon)-x_\nu(t)]\rangle}{\epsilon}=0,
\label{eqt1}
\end{equation}
where we have used the first correlator of eq.~(\ref{xcorrels}) for the 
final equality. Likewise, one may write
\begin{equation}
\langle \dot{x}_\mu(t)\dot{x}_\nu(t)\rangle=-\langle x_\mu(t)\ddot x_{\nu}(t)
\rangle
\label{eqt2}
\end{equation}
(using integration by parts), which gives
\begin{equation}
\langle x_\mu(t)\ddot x_{\nu}(t)
\rangle=-\lim_{\epsilon\rightarrow0^+}\frac{\langle x_\mu(t)[
x_\nu(t+2\epsilon)-x_\nu(t+\epsilon)+x_\nu(t)]\rangle}{\epsilon^2}=0.
\label{eqt3}
\end{equation}

We now consider the external line factor of eq.~(\ref{fT2}), where for brevity
in what follows we write $p_j\rightarrow p$. Furthermore, we may use the
symmetry of $h^{\mu\nu}$ to rewrite eq.~(\ref{fT2}) in the 
$T\rightarrow\infty$ limit as
\begin{align}
f(\infty)&=\int{\cal D}x\,\exp\left\{i\int_0^\infty dt\left[\frac{\dot{x}^2}{2}
-\frac{\kappa}{2}p_{(\nu}\dot{x}_{\mu)}h^{\mu\nu}-\frac{i\kappa}{2}
(\partial_\mu h^{\mu\nu})\dot{x}_\nu+\frac{\kappa^2}{8}\left(p_\nu \,p_\beta\,
\eta_{\alpha\mu}+p_{\mu}\,p_\beta\,\eta_{\alpha\nu}\right.\right.\right.
\notag\\
&\left.\left.\left.+p_\nu\,p_\alpha\,
\eta_{\beta\mu}+p_{\mu}\,p_\alpha\,\eta_{\beta\nu}\right)h^{\mu\nu}
h^{\alpha\beta}-\frac{\kappa}{2}p_{\mu}\,p_{\nu}h^{\mu\nu}-\frac{i\kappa}{4}
p_{(\nu}\partial_{\mu)}h^{\mu\nu}-\frac{m^2}{2}\frac{\kappa}{d-2}\eta_{\mu\nu}
h^{\mu\nu}\right.\right.\notag\\
&\left.\left.+\frac{i\kappa^2}{4}p_{(\nu}\eta_{\mu)\beta}h^{\mu\nu}
(\partial_\alpha h^{\alpha\beta})-\frac{m^2\kappa^2}{2}\left(
\frac{\eta_{\mu\nu}\,\eta_{\alpha\beta}}{(d-2)^2}-
\frac{(\eta_{\mu\beta}\,\eta_{\nu\alpha}+\eta_{\nu\beta}\eta_{\mu\alpha})}
{2(d-2)}\right)h^{\mu\nu}h^{\alpha\beta}
\right]\right\}
\label{fTpath1}
\end{align}
In this expression, we have anticipated that some of the terms in
eq.~(\ref{fT2}) will not lead to NE contributions after carrying out the 
path integral, and so have already discarded them (e.g. terms with both powers
of $x$ and more than one graviton tensor). Ignoring terms which depend 
explicitly on $m^2$ for the moment, one may expand the arguments of the
graviton tensors (each of which is evaluated at 4-position $p^\mu t+x^\mu$)
where relevant to give
\begin{align}
f(\infty)&=\int{\cal D}x\exp\left\{i\int_0^\infty dt\left[\frac{\dot{x}^2}{2}
-\frac{\kappa}{2}p_{(\nu}\dot{x}_{\mu)}h^{\mu\nu}-\frac{i\kappa}{2}
(\partial_\mu
h^{\mu\nu})\dot{x}_\nu+\frac{\kappa^2}{8}\left(p_\nu\,p_\beta\,
\eta_{\mu\alpha}+p_\mu\,p_\beta\,\eta_{\alpha\nu}\right.\right.\right.\notag\\
&\left.\left.\left.+p_\nu\,p_\alpha\,
\eta_{\beta\mu}+p_\mu\,p_\alpha\,\eta_{\beta\nu}\right)h^{\mu\nu}
h^{\alpha\beta}-\frac{\kappa}{2}p_\mu\,p_\nu\,h^{\mu\nu}-\frac{\kappa}{2}
p_{\mu}\,p_\nu(\partial_\sigma h^{\mu\nu})x^\sigma\right.\right.\notag\\
&\left.\left.-\frac{\kappa}{4}p_{\mu}
\,p_{\nu}(\partial_\sigma\partial_\tau h^{\mu\nu})x^\sigma x^\tau
-\frac{i\kappa}{4}p_{(\nu}\partial_{\mu)}h^{\mu\nu}-\frac{i\kappa}{4}
p_{(\nu}\partial_{\mu)}(\partial_\sigma h^{\mu\nu})x^\sigma\right.\right.\\
&\left.\left.
+\frac{i\kappa^2}{4}p_{(\nu}\,\eta_{\mu)\beta}h^{\mu\nu}(\partial_\alpha
h^{\alpha\beta})
\right]\right\},
\label{fTpath2}
\end{align}
where all graviton tensors are now evaluated on the classical trajectory.
Also, we have neglected terms of form $x^{\sigma}\dot{x}^{\tau}$ and
$\dot{x}^{\sigma}\dot{x}^{\tau}$, which would vanish in the following
calculation due to the equal time commutators of eqs.~(\ref{eqt2})
and~(\ref{eqt3}). It is now convenient to transform to momentum space
(in order to ultimately be able to read off the momentum space Feynman rules).
Writing
\begin{equation}
h^{\mu\nu}(x)=\int\frac{d^d k}{(2\pi)^d}h^{\mu\nu}(k)e^{ik\cdot p\,t},
\label{Fourier}
\end{equation}
(where as in the rest of the paper we use the same symbol for the Fourier
transformed graviton field), the graviton contributions to the exponent 
in eq.~(\ref{fTpath2}) can be written
\begin{align}
i\int_0^\infty &dt\left\{\int\frac{d^dk}{(2\pi)^d}h^{\mu\nu}(k)
e^{ip\cdot k\,t}\left(-\frac{\kappa}{2}p_{(\nu}\dot{x}_{\mu)}+\frac{\kappa}
{4}k_{(\mu}\dot{x}_{\nu)}-\frac{\kappa}{2}p_\mu\,p_\nu-\frac{i\kappa}{2}
p_\mu\,p_\nu\,k^\sigma\, x_\sigma\right.\right.\notag\\
&\left.\left.\quad+\frac{\kappa}{4}p_{\mu}p_\nu\,k^\sigma\,
k^\tau\,x_{\sigma}\,x_\tau+\frac{\kappa}{4}p_{(\nu}k_{\mu)}
+\frac{i\kappa}{4}p_{(\nu}k_{\mu)}k^\sigma\,x_\sigma\right)\right.\notag\\
&\left.
+\int\frac{d^dk}{(2\pi)^d}\int\frac{d^dl}{(2\pi)^d}h^{\mu\nu}(k)
h^{\alpha\beta}(l)e^{ip\cdot(k+l)t}\left[-\frac{\kappa^2}{4}
p_{(\nu}\eta_{\mu)\beta}k_\alpha+\frac{\kappa^2}{8}\left(
p_\nu\,p_\beta\,\eta_{\alpha\mu}+p_\mu\,p_\beta\,\eta_{\alpha\nu}\right.\right.
\right.\notag\\
&\left.\left.\left.\quad
+p_\nu\,p_\alpha\,\eta_{\beta\mu}+p_\mu\,p_\alpha\,\eta_{\beta\nu}\right)
\phantom{\int}\right]\right\}.
\label{fTpath3}
\end{align}
In this expression, the terms with no $x$ powers are
\begin{align}
&i\int_0^\infty e^{ip\cdot k\,t}\left[-\frac{\kappa}{2}p_\mu\,p_\nu
+\frac{\kappa}{4}p_{(\nu}k_{\mu)}-\frac{\kappa^2}{4}e^{ip\cdot l\,t}
p_{(\nu}\eta_{\mu)\beta}k_\alpha+\frac{\kappa^2}{8}e^{ip\cdot l\,t}
\left(p_\nu\,p_\beta\,\eta_{\alpha\mu}+p_\mu\,p_\beta\,\eta_{\alpha\nu}
\right.\right.\notag\\
&\left.\quad\left.
+p_\nu\,p_\alpha\,\eta_{\beta\mu}+p_\mu\,p_\alpha\,\eta_{\beta\nu}\right)
\phantom{\int}\right],
\label{0xterms}
\end{align}
where to shorten the notation (and from now on) we omit 
the integrals over momenta
and the Fourier transformed graviton fields, which can easily be
reinstated based on the number of Lorentz indices in each term. 
The various terms in eq.~(\ref{0xterms}) can
immediately be interpreted in terms of Feynman rules, as the remaining path
integral over $x$ gives unity, assuming the appropriate normalization.
Terms with one power of $x^\mu$ generate vertices for the coordinate $x^{\mu}$,
which we label as follows:
\begin{align}
i\int_0^\infty dt e^{ip\cdot k\,t}\left[\overbrace{-\frac{\kappa}{2}
p_{(\nu}\dot{x}_{\mu)}}^{(A)}\underbrace{+\frac{\kappa}{4}k_{(\mu}
\dot{x}_{\nu)}}_{(B)}\overbrace{-\frac{i\kappa}{2}p_\mu\,p_\nu\,k^\sigma\,
x_\sigma}^{(C)}\underbrace{+\frac{i\kappa}{4}p_{(\nu}k_{\mu)}k^\sigma\,
x_\sigma}_{(D)}\right].
\label{1xterms}
\end{align}
In carrying out the path integral over $x$ to NE order, one must construct 
all tree level graphs containing two such vertices. Again omitting the
graviton fields and momentum integrals, we start by considering the 
(A)--(A) combination, which tree level graph gives
\begin{align}
&\frac{1}{2}\left(-\frac{i\kappa}{2}\right)^2\int_0^\infty dt\int_0^\infty dt'
\,e^{i(p\cdot kt+p\cdot lt')}\langle [p_\nu\,\dot{x}_\mu(t)+p_\mu
\dot{x}_\nu(t)]
[p_\beta\,\dot{x}_\alpha(t')+p_\alpha\dot{x}_\beta(t')]\rangle,
\label{AA1}
\end{align}
where the initial factor of $1/2$ is a symmetry factor. Inserting the 
correlator of eq.~(\ref{xcorrels}), eq.~(\ref{AA1}) becomes
\begin{equation}
-\frac{i\kappa^2}{8}\int_0^\infty dt\,e^{ip\cdot(k+l)t}\left[
p_\nu\,p_\beta\,\eta_{\alpha\mu}+p_\mu\,p_\beta\,\eta_{\alpha\nu}
+p_\nu\,p_\alpha\,\eta_{\mu\beta}+p_\mu\,p_\alpha\,\eta_{\beta\nu}
\right].
\label{AA2}
\end{equation}
Note that this precisely cancels the contribution from the final term in
eq.~(\ref{0xterms}), which thus does not result in a contribution to
the effective Feynman rules at NE order. 

Next, one has the (A)--(B) combination, which gives 
\begin{align}
&\left(-\frac{i\kappa}{2}\right)\left(\frac{i\kappa}{4}\right)
\int_0^\infty dt\int_0^\infty dt'\,e^{i(p\cdot kt+p\cdot lt')}
\langle[p_\nu\,\dot{x}_\mu(t)+p_\mu\,\dot{x}_\nu(t)]
[k_\alpha\dot{x}_\beta(t')+k_\beta\dot{x}_\alpha(t')]\rangle\notag\\
&=\frac{i\kappa^2}{8}\int_0^\infty dt\,e^{ip\cdot(k+l)t}
\left[p_\nu\,k_\alpha\,\eta_{\mu\beta}+p_\mu\,k_\alpha\,\eta_{\nu\beta}
+p_\nu k_\beta\,\eta_{\mu\alpha}+p_\mu\,k_\beta\,\eta_{\nu\alpha}
\right].
\label{AB1}
\end{align}
Exploiting the fact that there is an implicit factor of
$h^{\mu\nu}(k)h^{\alpha\beta}(l)$ in eq.~(\ref{0xterms}), 
we may further rewrite eq.~(\ref{AB1}) as
\begin{equation}
\frac{i\kappa^2}{16}\int_0^\infty dt\,e^{ip\cdot(k+l)t}
\left[p_\nu\,k_\alpha\,\eta_{\mu\beta}+p_\mu\,k_\alpha\,\eta_{\nu\beta}
+p_\nu k_\beta\,\eta_{\mu\alpha}+p_\mu\,k_\beta\,\eta_{\nu\alpha}
+(k\leftrightarrow l)\right].
\label{AB1b}
\end{equation}
By the same token, the third term in eq.~(\ref{0xterms}) can be written
\begin{equation}
-\frac{i\kappa^2}{16}\int_0^\infty dt\,e^{ip\cdot(k+l)t}\left[p_\nu\,
k_\alpha\,\eta_{\mu\beta}+p_\mu\,k_\alpha\,\eta_{\nu\beta}
+p_\nu\,k_\beta\,\eta_{\mu\alpha}+p_\mu\,k_\beta\,\eta_{\nu\alpha}
+(\mu,\nu,k\leftrightarrow\alpha,\beta,l)\right],
\label{AB2}
\end{equation}
The terms shown in eq.~(\ref{AB2}) precisely cancel eq.~(\ref{AB1b}), and 
thus there is again no
contribution to the effective Feynman rules.

Next up is the combination (A)--(C), which gives
\begin{align}
&\left(-\frac{i\kappa}{2}\right)\left(\frac{\kappa}{2}\right)
p_\alpha\,p_\beta\,k^\sigma\int_0^\infty dt\int_0^\infty dt'
e^{i(p\cdot kt+ip\cdot lt')}\langle[p_\nu\,\dot{x}_\mu(t)+p_\mu\dot{x}_\nu(t)]
x_\sigma(t')\notag\\
&=\frac{\kappa^2}{4}p_\alpha\,p_\beta\,p_{(\mu}k_{\nu)}\int_0^\infty dt
\int_0^\infty dt'\Theta(t-t')e^{i(p\cdot kt+p\cdot lt')}\notag\\
&=-\frac{\kappa^2}{4}\frac{p_\alpha\,p_\beta\,p_{(\mu}k_{\nu)}}{p\cdot k\,
p\cdot(k+l)}\notag\\
&\equiv -\frac{\kappa^2}{16}\left[\frac{p_\alpha\,
p_\beta\,p_{(\mu}k_{\nu)}}{p\cdot k\,p\cdot(k+l)}+\frac{p_\alpha\,
p_\beta\,p_{(\mu}l_{\nu)}}{p\cdot l\,
p\cdot(k+l)}+(\mu\nu\leftrightarrow\alpha\beta)\right],
\label{AC1}
\end{align}
where have symmetrized over graviton indices in the final line.
This is indeed a NE contribution, and thus contributes to the effective
two-graviton vertex. 

By a similar procedure, one may evaluate the contributions from all other
combinations of the vertices in eq.~(\ref{1xterms}). The reader may verify
that (A)--(D), (B)--(B), (B)--(C), (B)--(D), (C)--(D) and (D)--(D) do not
give any contribution up to NE order. For (C)--(C), one has 
\begin{align}
&\frac{1}{2}\left(\frac{\kappa}{2}\right)^2p_\mu\,p_\nu\,p_\alpha\,p_\beta
\,k^\sigma\,l^\tau\int_0^\infty dt\int_0^\infty dt'x_\sigma(t)x_\tau(t')
e^{i(p\cdot kt+p\cdot lt')}\notag\\
&=+\frac{i\kappa^2}{8}p_\mu\,p_\nu\,p_\alpha\,p_\beta\,k\cdot l\int_0^\infty
dt\int_0^\infty dt'\,\min(t,t')e^{i(p\cdot kt+ip\cdot lt')}\notag\\
&=\frac{\kappa^2}{8}\frac{p_\mu\,p_\nu\,p_\alpha\,p_\beta\,k\cdot l}
{p\cdot k\,p\cdot l\,p\cdot(k+l)}.
\label{CC1}
\end{align}
This is already symmetric under $(\mu\nu\leftarrow\alpha\beta)$.

One must also consider the term in eq.~(\ref{fTpath3}) with two powers of
$x$. This contributes to a loop graph at NE order, which gives a contribution 
(including a symmetry factor of $1/2$)
\begin{align}
&\frac{1}{2}\frac{i\kappa}{4}p_\mu\,p_\nu\,k^\sigma\,k^\tau\int_0^\infty dt
x_\sigma(t)x_\tau(t)e^{ip\cdot kt}\notag\\
&=-\frac{\kappa}{8}p_\mu\,p_\nu\,k^2\int_0^\infty dt\,te^{ip\cdot kt}\notag\\
&=\frac{\kappa}{8}\frac{p_\mu\,p_\nu\,k^2}{(p\cdot k)^2}.
\label{twox1}
\end{align}
This is already symmetric in $\mu$ and $\nu$. Furthermore, the fact that this
term is quadratic in $x$ means that the symmetry factor of $1/2$ does not enter
the resulting Feynman rule. 

The uncancelled terms with no $x$ powers from eq.~(\ref{0xterms}) give
\begin{align}
-\frac{i\kappa}{2}p_\mu\,p_\nu\int_0^\infty dt\,e^{ip\cdot kt}=\frac{\kappa}{2}
\frac{p_\mu\,p_\nu}{p\cdot k};\notag\\
\frac{i\kappa}{4}p_{(\mu}k_{\nu)}\int_0^\infty dt e^{ip\cdot kt}=
-\frac{\kappa}{4}\frac{p_{(\mu}\,k_{\nu)}}{p\cdot k}.
\label{0xtermsrules}
\end{align}

Above we have ignored $m^2$ dependent terms, and it is straightforward to
repeat the above analysis for these. Expanding arguments of graviton
tensors where necessary and keeping only relevant terms, the mass-related 
terms in eq.~(\ref{fTpath1}) give
\begin{align}
&i\int_0^\infty dt\left[-\frac{m^2}{2}\frac{\kappa}{d-2}\eta_{\mu\nu}h^{\mu\nu}
-\frac{m^2}{2}\frac{\kappa}{d-2}\eta_{\mu\nu}(\partial_\sigma h^{\mu\nu})
x^\sigma-\frac{m^2}{4}\frac{\kappa}{d-2}\eta_{\mu\nu}(\partial_\sigma
\partial_\tau h^{\mu\nu})x^\sigma x^\tau\right.\notag\\
&\left.\quad-\frac{m^2\kappa^2}{2}
\left(\frac{\eta_{\mu\nu}\,\eta_{\alpha\beta}}{(d-2)^2}
-\frac{(\eta_{\mu\beta}\,\eta_{\nu\alpha}+\eta_{\nu\beta}\,
\eta_{\mu\alpha})}{2(d-2)}\right)h^{\mu\nu}h^{\alpha\beta}\right],
\label{mterms}
\end{align}
which in momentum space becomes
\begin{align}
\int_0^\infty&\left\{\int\frac{d^dk}{(2\pi)^d}h^{\mu\nu}(k)
e^{ip\cdot kt}\left[-\frac{im^2}{2}\frac{\kappa}{d-2}\eta_{\mu\nu}
+\frac{m^2}{2}\frac{\kappa}{d-2}\eta_{\mu\nu}\,k^\sigma\,x_\sigma
+\frac{im^2}{4}\frac{\kappa}{d-2}\eta_{\mu\nu}\,k^\sigma\,k^\tau\,
x_\sigma\,x_\tau\right]\right.\notag\\
&\left.+\int\frac{d^dk}{(2\pi)^d}\frac{d^dl}{(2\pi)^d}
e^{ip\cdot(k+l)t}\left[-\frac{im^2\kappa^2}{2}\frac{\eta_{\mu\nu}\,
\eta_{\alpha\beta}}{(d-2)^2}+\frac{im^2\kappa^2}{4}\frac{
(\eta_{\mu\beta}\,\eta_{\nu\alpha}+\eta_{\nu\beta}\,\eta_{\mu\alpha})}
{d-2}\right]h^{\mu\nu}(k)h^{\alpha\beta}(l)\right\}.
\label{mtermsmom}
\end{align}
The terms with no powers of $x$ give (again omitting the graviton fields
and momentum integrals for brevity)
\begin{align}
&-i\int_0^\infty dt\,e^{ip\cdot kt}\frac{m^2}{2}\frac{\kappa}{d-2}
\eta_{\mu\nu}=\frac{m^2}{2}\frac{\kappa}{d-2}\frac{\eta_{\mu\nu}}{p\cdot k};
\label{m0termsa}\\
&\int_0^\infty dt\,e^{ip\cdot(k+l)t}\left(-\frac{im^2\kappa^2}{2}\right)
\left[\frac{\eta_{\mu\nu}\,\eta_{\alpha\beta}}{(d-2)^2}
-\frac{1}{2(d-2)}(\eta_{\mu\beta}\,\eta_{\nu\alpha}+\eta_{\nu\beta}\,
\eta_{\mu\alpha})\right]\notag\\
&=\frac{m^2\kappa^2}{4p\cdot(k+l)}\left[\frac{2\eta_{\mu\nu}\,
\eta_{\alpha\beta}}{(d-2)^2}
-\frac{1}{(d-2)}(\eta_{\mu\beta}\,\eta_{\nu\alpha}+\eta_{\nu\beta}\,
\eta_{\mu\alpha})\right].
\label{m0terms}
\end{align}
The term in eq.~(\ref{mtermsmom}) which has one power of $x$ gives a 
vertex for $x^\mu$
\begin{equation}
\frac{m^2}{2}\frac{\kappa}{d-2}\int_0^\infty e^{ip\cdot kt}\eta_{\mu\nu}
\,k^\sigma x_\sigma.
\label{m1xterm}
\end{equation}
Labelling this by (E), one must consider all tree level graphs in which 
(E) is contracted with one of the vertices (A) to (D) of eq.~(\ref{1xterms}).
First of all, (E)--(A) gives
\begin{align}
&\frac{m^2}{2}\frac{\kappa}{d-2}\left(-\frac{i\kappa}{2}\right)
\eta_{\mu\nu}\,k^\sigma\int_0^\infty dt\int_0^\infty dt'\langle
[p_\alpha\,\dot{x}_\beta(t')+p_\beta\,\dot{x}_\alpha]x_\sigma(t)
e^{i(p\cdot kt+p\cdot lt')}\rangle\notag\\
&=-\frac{m^2\kappa^2}{4(d-2)}\frac{\eta_{\mu\nu}\,p_{(\alpha}
k_{\beta)}}{p\cdot k\,p\cdot(k+l)}\notag\\
&\equiv -\frac{m^2\kappa^2}{16(d-2)}\left[\frac{\eta_{\mu\nu}\,p_{(\alpha}
k_{\beta)}}{p\cdot k\,p\cdot(k+l)}+\frac{\eta_{\mu\nu}\,p_{(\alpha}
l_{\beta)}}{p\cdot l\,p\cdot(k+l)}+(\mu\nu\leftrightarrow\alpha\beta)\right].
\label{EA1}
\end{align}
The reader may check that (E)--(B) and (E)--(D) do not
give NE contributions. The combination (E)--(C) gives
\begin{align}
&\frac{m^2}{2}\frac{\kappa}{d-2}\eta_{\mu\nu}\,k^\sigma\left(\frac{\kappa}{2}
p_\alpha\,p_\beta\,l^\tau\right)\int_0^\infty dt\int_0^\infty dt'\,x_\sigma\,
x_\tau\,e^{i(p\cdot kt+p\cdot lt')}\notag\\
&=\frac{m^2}{4}\frac{\kappa^2}{d-2}\frac{\eta_{\mu\nu}\,p_\alpha\,p_\beta\,
k\cdot l}{p\cdot k\,p\cdot l\,p\cdot(k+l)}\notag\\
&\equiv\frac{m^2}{8}\frac{\kappa^2}{d-2}\left[\frac{\eta_{\mu\nu}\,
p_\alpha\,p_\beta\,k\cdot l}{p\cdot k\,p\cdot l\,p\cdot(k+l)}
+(\mu\nu\leftrightarrow\alpha\beta)\right].
\label{EC1}
\end{align}

Finally one has (E)--(E) which gives 
\begin{align}
&\frac{1}{2}\left(\frac{m^2}{2}\frac{\kappa}{d-2}\right)^2\eta_{\mu\nu}
\,\eta_{\alpha\beta}\,k^\sigma\,l^\tau\int_0^\infty dt\int_0^\infty
dt'\,\langle x_\sigma(t)x_\tau(t')\rangle e^{i(p\cdot kt+p\cdot lt')}\notag\\
&=\frac{m^4\kappa^2}{8(d-2)}\frac{\eta_{\mu\nu}\,\eta_{\alpha\beta}\,
k\cdot l}{p\cdot k\,p\cdot l\,p\cdot(k+l)}.
\label{EE1}
\end{align}
There is also a term in eq.~(\ref{mtermsmom}) which has two powers of $x$.
This gives a loop contribution to the path integral
\begin{equation}
\frac{1}{2}\frac{im^2}{4}\frac{\kappa}{d-2}\eta_{\mu\nu}\,k^\sigma\,k^\tau
\int_0^\infty dt e^{ip\cdot kt}\langle x_\sigma(t)\,x_\tau(t)\rangle=
\frac{m^2}{8}\frac{\kappa}{d-2}\eta_{\mu\nu}\frac{k^2}{(p\cdot k)^2},
\label{m2xterm}
\end{equation}
where again the symmetry factor of $1/2$ does not enter the Feynman rule due
to the fact that this term is quadratic in $x$.

This completes the calculation of all contributions to the $x$ path integral
up to NE order. Collecting the results of eqs.~(\ref{AC1}, \ref{CC1}, 
\ref{twox1}, \ref{0xtermsrules}, \ref{m0termsa}, \ref{m0terms}, \ref{EA1}, \ref{EC1}, 
\ref{EE1}, \ref{m2xterm}), one indeed finds that the external line factor 
is given by eq.~(\ref{expmom}).

\section{Diagrammatic check of the NE Feynman rules}
\label{app:diags}
In this appendix, we cross-check the effective Feynman rules for graviton
emission up to next-to-eikonal order, using a diagrammatic approach.
Our analysis is similar to that presented in appendix B 
of~\cite{Laenen:2008gt}, and rests upon the fact that the effective
Feynman rules must reproduce the known NE limit of an exact calculation
which includes up to two graviton emissions.

\begin{figure}
\begin{center}
\scalebox{0.8}{\includegraphics{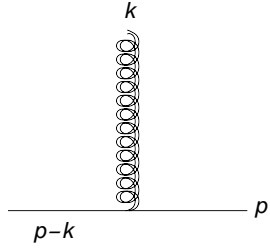}}
\caption{Graph used to check the one graviton 
emission vertex.}
\label{1gcheck}
\end{center}
\end{figure}

We begin by considering figure~\ref{1gcheck}, which contains a single graviton
emission from an external line of momentum $p$. 
The exact Feynman rule for the graviton emission is given in 
eq.~(\ref{feynrule2}), where $p_1$ and $p_2$ are the momenta to 
the left- and right-hand side of the vertex. Given that this assumes 
all momenta are outgoing\footnote{See also footnote 6 on p. 7.}, 
one must set $p_1=-(p-k)$ and $p_2=p$. Combined
with the propagator for the internal graviton line, this gives a total
contribution
\begin{equation}
-\frac{\kappa}{2}\frac{1}{(p-k)^2+m^2}\left((p-k)_{(\mu}p_{\nu)}+
\frac{2m^2}{d-2}\eta_{\mu\nu}\right).
\label{1ga}
\end{equation}
Expanding this up to NE order gives
\begin{align}
\frac{\kappa}{2}\frac{p_\mu\,p_\nu}{p\cdot k}-\frac{\kappa}{4}
\frac{p_{(\mu}k_{\nu)}}{p\cdot k}+\frac{m^2}{2}\frac{\kappa}{d-2}
\frac{\eta_{\mu\nu}}{p\cdot k}+\frac{\kappa}{4}\frac{p_\mu\,p_\nu\,
k^2}{(p\cdot k)^2}+\frac{m^2}{4}\frac{\kappa}{d-2}\eta_{\mu\nu}
\frac{k^2}{(p\cdot k)^2}.
\label{1gdiagterms}
\end{align}
This agrees with the results of eqs.~(\ref{1gErule}) and~(\ref{1gNErule}). 

\begin{figure}
\begin{center}
\scalebox{0.8}{\includegraphics{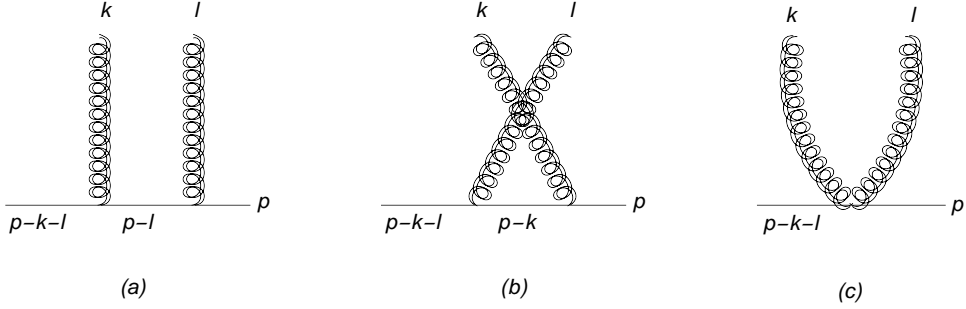}}
\caption{Graphs used to check the two graviton 
emission vertex.}
\label{2gcheck}
\end{center}
\end{figure}

To check the two-graviton vertex, we consider the graphs shown in
figure~\ref{2gcheck}. Firstly, there is the graph of figure~\ref{2gcheck}(a),
consisting of two separate single graviton emissions. 
This gives a contribution
\begin{align}
\frac{\kappa^2}{4}\left[\frac{(p-k-l)_{(\mu}(p-l)_{\nu)}}{(p-k-l)^2+m^2}
+\frac{2m^2}{d-2}\frac{\eta_{\mu\nu}}{(p-k-l)^2+m^2}\right]
\left[\frac{(p-l)_{(\alpha}p_{\beta)}}{(p-l)^2+m^2}+\frac{2m^2}{d-2}
\frac{\eta_{\alpha\beta}}{(p-l)^2+m^2}\right].
\label{2gcheck1}
\end{align}
In checking the two graviton vertex, one may combine this graph with its
counterpart in which the gluon momenta are permuted, with a factor $1/2$
to correct for the double counting. The sum should then separate into a 
part which factorizes into products of lower order emissions, and a 
non-factorizable remainder, which should agree with the two-graviton vertex
obtained using the path integral approach~\cite{Laenen:2008gt,Laenen:2010uz}.
The permuted graph is shown in figure~\ref{2gcheck}(b), and its contribution
can be easily obtained from eq.~(\ref{2gcheck1}) by interchanging
$k$ and $l$. 

There are two types of contribution in expanding the weighted sum 
of these graphs to NE order. Firstly,
there are terms which arise from expanding to ${\cal O}(k)$ or ${\cal O}(l)$
in the numerator. Secondly, there are terms which result from expanding the
denominators about their eikonal limit i.e.
\begin{equation}
\frac{1}{(p-k)^2+m^2}=-\frac{1}{2p\cdot k}\left[1+\frac{k^2}{2p\cdot k}+\ldots
\right],
\label{denexp}
\end{equation}
where the ellipsis denotes terms beyond the NE approximation. Considering
first the numerator contributions, expanding eq.~(\ref{2gcheck1}) and combining
this with figure~\ref{2gcheck}(b) gives
\begin{align}
&\frac{\kappa^2}{8}\left\{\left[\frac{p_\mu\,p_\nu\,p_\alpha\,p_\beta}
{p\cdot(k+l)}-\frac{(k+l)_{(\mu}p_{\nu)}\,p_\alpha\,p_\beta}{2p\cdot(k+l)}
+\frac{m^2}{d-2}\frac{(\eta_{\mu\nu}\,p_\alpha\,p_\beta
+\eta_{\alpha\beta}\,p_\mu\,p_\nu)}{p\cdot(k+l)}\right.\right.\notag\\
&\left.\left.-\frac{m^2}{2(d-2)}
\frac{\eta_{\alpha\beta}(k+l)_{(\mu}p_{\nu)}}{p\cdot(k+l)}
+\frac{4m^4}{(d-2)^2}\frac{\eta_{\mu\nu}\,\eta_{\alpha\beta}}{p\cdot(k+l)}
\right]
\left(\frac{1}{p\cdot k}+\frac{1}{p\cdot l}\right)
-\frac{l_{(\mu}p_{\nu)}\,p_\alpha\,p_\beta}{2p\cdot(k+l)\,p\cdot l}
\right.\notag\\
&\left.-\frac{k_{(\mu}p_{\nu)}\,p_\alpha\,p_\beta}{2p\cdot(k+l)\,p\cdot k}
-\frac{l_{(\alpha}p_{\beta)}\,p_\mu\,p_\nu}{2p\cdot(k+l)\,p\cdot l}
-\frac{k_{(\alpha}p_{\beta)}\,p_\mu\,p_\nu}{2p\cdot(k+l)\,p\cdot k}
-\frac{m^2}{2(d-2)}\frac{\eta_{\alpha\beta}\,l_{(\mu}p_{\nu)}}
{p\cdot l\,p\cdot(k+l)}\right.\notag\\
&\left.-\frac{m^2}{2(d-2)}\frac{\eta_{\alpha\beta}\,k_{(\mu}p_{\nu)}}
{p\cdot k\,p\cdot(k+l)}
-\frac{m^2}{2(d-2)}\frac{\eta_{\mu\nu}\,k_{(\alpha}p_{\beta)}}
{p\cdot k\,p\cdot(k+l)}-\frac{m^2}{2(d-2)}\frac{\eta_{\mu\nu}\,l_{(\alpha}
p_{\beta)}}{p\cdot l\,p\cdot(k+l)}
\right\}.
\label{2gcheck2}
\end{align}
The square-bracketed terms factorize upon application of the eikonal identity
\begin{equation}
\frac{1}{p\cdot(k+l)}\left(\frac{1}{p\cdot k}+\frac{1}{p\cdot l}\right)
=\frac{1}{p\cdot k\,p\cdot l},
\label{eikid}
\end{equation}
to give products of lower order effective Feynman rules. Left are the
non-factorizable contributions
\begin{align}
-\frac{\kappa^2}{16}&\left[\frac{l_{(\mu}p_{\nu)}\,p_\alpha\,p_\beta}
{p\cdot(k+l)\,p\cdot l}+\frac{k_{(\mu}p_{\nu)}\,p_\alpha\,p_\beta}
{p\cdot(k+l)\,p\cdot k}+\frac{m^2}{d-2}\frac{\eta_{\alpha\beta}\,
l_{(\mu}p_{\nu)}}{p\cdot(k+l)\,p\cdot l}+\frac{m^2}{d-2}\frac{
\eta_{\alpha\beta}\,k_{(\mu}p_{\nu)}}{p\cdot(k+l)\,p\cdot k}
\right.\notag\\
&\left.+(\mu\nu\leftrightarrow\alpha\beta)\phantom{\Big{|}}\right].
\label{nonfac1}
\end{align}
These thus contribute to the effective two-graviton vertex.

Now considering the denominator terms in eq.~(\ref{2gcheck1}) (together
with the corresponding result with $k\leftrightarrow l$), the NE terms
from the $m^2$-independent terms are
\begin{align}
\frac{\kappa^2}{8}\frac{p_\mu\,p_\nu\,p_\alpha\,p_\beta}
{p\cdot(k+l)}\left\{\frac{(k+l)^2}{2p\cdot(k+l)}\left(\frac{1}{p\cdot k}
+\frac{1}{p\cdot l}\right)+\frac{k^2}{2(p\cdot k)^2}+\frac{l^2}{2(p\cdot l)^2}
\right\},
\end{align}
which can be rearranged to give
\begin{equation}
+\frac{\kappa^2}{16}\frac{p_\mu\,p_\nu\,p_\alpha\,p_\beta}{p\cdot l}
\frac{k^2}{(p\cdot k)^2}+\frac{\kappa^2}{16}\frac{p_\mu\,p_\nu\,p_\alpha\,
p_\beta}{p\cdot k}\frac{l^2}{(p\cdot l)^2}+\frac{\kappa^2}{8}\frac{p_\mu\,
p_\nu\,p_\alpha\,p_\beta\,k\cdot l}{p\cdot k\,p\cdot l\,p\cdot(k+l)}.
\label{den1}
\end{equation}
The first two terms have factorized into products of lower order effective 
Feynman rules. The final term, however, is non-factorizable and hence 
contributes to the effective two-graviton vertex. 

Now considering the $m^2$-dependent denominator contributions, these are
given by (keeping only NE terms)
\begin{align}
\frac{\kappa^2}{8}&\left\{\left[\frac{m^2}{d-2}\frac{(\eta_{\mu\nu}
\,p_{\alpha}\,p_{\beta}
+\eta_{\alpha\beta}\,p_\mu\,p_\nu)}{p\cdot(k+l)p\cdot l}+\frac{m^4}
{(d-2)^2}\frac{\eta_{\mu\nu}\,\eta_{\alpha\beta}}{p\cdot(k+l)\,p\cdot l}
\right]\left(\frac{(k+l)^2}{2p\cdot(k+l)}+\frac{l^2}{2p\cdot l}\right)
\right.\notag\\
&\left.+(k\leftrightarrow l)\phantom{\Big{|}}\right\}.
\label{denm1}
\end{align}
Rearranging and keeping non-factorizable terms only, one finds
\begin{equation}
\frac{\kappa^2m^2}{8(d-2)}\frac{(\eta_{\mu\nu}\,p_\alpha\,p_\beta
+\eta_{\alpha\beta}\,p_\mu\,p_\nu) k\cdot l}
{p\cdot k\,p\cdot l\,p\cdot(k+l)}+\frac{\kappa^2m^4}{8(d-2)^2}
\frac{\eta_{\mu\nu}\,\eta_{\alpha\beta}\,k\cdot l}{p\cdot k\,p\cdot l\,
p\cdot(k+l)}.
\label{denm2}
\end{equation}

Finally, there is the graph of figure~\ref{2gcheck}(c), which is given by
the exact two-graviton vertex~\cite{Hamber:2007fk}
\begin{equation}
\frac{\kappa^2m^2}{2(d-2)}\left[\eta_{\mu\alpha}\,\eta_{\nu\beta}
+\eta_{\mu\beta}\,\eta_{\nu\alpha}-\frac{2}{d-2}\eta_{\mu\nu}\,
\eta_{\alpha\beta}\right].
\label{2gexact}
\end{equation}
combined with an eikonal propagator, to give
\begin{equation}
-\frac{\kappa^2m^2}{4(d-2)}\frac{1}{p\cdot (k+l)}\left[\eta_{\mu\alpha}\,\eta_{\nu\beta}
+\eta_{\mu\beta}\,\eta_{\nu\alpha}-\frac{2}{d-2}\eta_{\mu\nu}\,
\eta_{\alpha\beta}\right].
\label{2gexact2}
\end{equation}

Adding this to the non-factorizable contributions from 
eqs.~(\ref{nonfac1}, \ref{den1}, \ref{denm2}), one finds a total
effective two-graviton vertex at NE order that is equal to that presented
in eq~(\ref{2gNErule}) (obtained from the path integral approach).
One could of course have used the diagrammatic
method of this appendix to derive the NE vertices in the first place, without
relying on the path integral approach. However, there is then no proof that
the effective Feynman rules hold to all orders in the perturbation expansion
i.e. that they are genuine Feynman rules\footnote{An all order diagrammatic
proof of the NE Feynman rules in QED and QCD has been given 
in~\cite{Laenen:2010uz}.}. This latter point is made manifest
by the path integral method. 

\section{Diagrammatic derivation of eq.~(\ref{gravlow}) for internal emissions}
\label{app:low}
In this appendix, we give an example of how eq.~(\ref{gravlow}) (which 
expresses how the subamplitude for soft graviton emission from within the 
hard interaction can be related to the hard interaction with no additional
emission) can be derived from a traditional Feynman diagram calculation.

\begin{figure}
\begin{center}
\scalebox{1.0}{\includegraphics{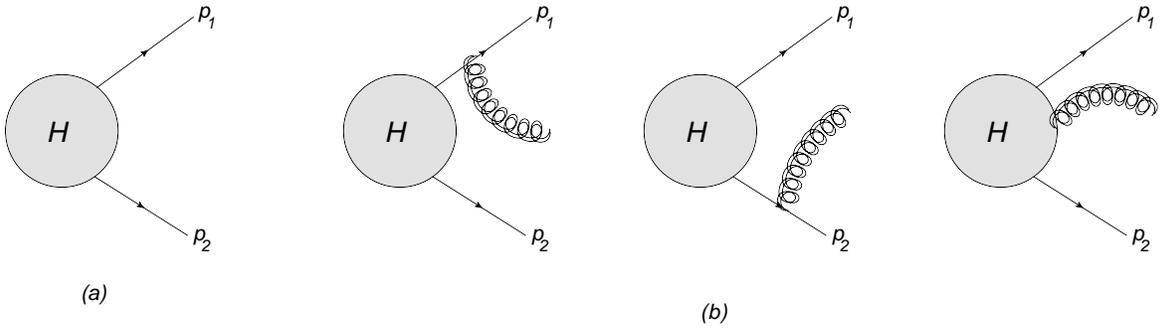}}
\caption{(a) Example vertex function involving two outgoing particles; 
(b) Series of diagrams involving all possible placements of an extra graviton.}
\label{vertexfig}
\end{center}
\end{figure}

We start with the vertex function shown in figure~\ref{vertexfig}(a) with two
external scalar particles, which in figure~\ref{vertexfig}(b) is dressed by an 
additional graviton emission in all possible places. 
For simplicity, we deal with the case of massless external particles. 
The vertex with no emission can then only depend on the invariant
$p_1\cdot p_2$, and we denote this by $\Gamma(p_1,p_2)$. The full Feynman rule
for emission of a graviton from a scalar is given by~\cite{Hamber:2007fk} 
\begin{equation}
\frac{\kappa}{2}\left(p_{1\mu}\,p_{2\nu}+p_{1\nu}\,p_{2\mu}-\frac{2}{d-2}m^2
\eta_{\mu\nu}\right)
\label{feynrule2}
\end{equation}
in the weak field approximation of eq.~(\ref{hdef}). For $m=0$ the vertex
function including an extra graviton emission (from figure~\ref{vertexfig}(b))
is then
\begin{align}
\Gamma_{\mu\nu}&=\frac{\kappa}{2}\left[\frac{p_{1\mu}(p_{1}-k)_\nu
+(p_1-k)_\mu p_{1\nu}}{-2p_1\cdot k}\Gamma[(p_1-k)\cdot p_2]\right.\notag\\
&\left.\quad
+\frac{p_{2\mu}(p_2-k)_\nu+(p_2-k)_\mu p_{2\nu}}{-2p_2\cdot k}\Gamma[p_1
\cdot(p_2-k)]\right]+\Gamma^{\rm int.}_{\mu\nu},
\label{gam1}
\end{align}
where $\Gamma^{\rm int.}_{\mu\nu}$ represents the contribution from the
internal emission graph. It will be convenient to symmetrize on both sides,
defining $\bar{\Gamma}_{\mu\nu}=\Gamma_{\mu\nu}+\Gamma_{\nu\mu}$ (and likewise
for $\Gamma^{\rm int.}_{\mu\nu}$). Then eq.~(\ref{gam1}) can be rewritten
\begin{align}
\frac{1}{2}\bar{\Gamma}_{\mu\nu}&=\frac{\kappa}{2}\left[
\frac{p_{1\mu}(p_{1}-k)_\nu+(p_1-k)_\mu p_{1\nu}}{-2p_1\cdot k}
\Gamma[(p_1-k)\cdot p_2]\right.\notag\\
&\left.\quad+\frac{p_{2\mu}(p_2-k)_\nu+(p_2-k)_\mu p_{2\nu}}{-2p_2\cdot k}
\Gamma[p_1\cdot(p_2-k)]\right]+\frac{1}{2}\bar{\Gamma}^{\rm int.}_{\mu\nu},
\label{gam2}
\end{align}
Expanding this to first order in $k$ gives
\begin{align}
\frac{1}{2}\bar{\Gamma}_{\mu\nu}&=\frac{\kappa}{2}\left[\frac{p_{1\mu}p_{1\nu}}
{-p_1\cdot k}\left(\Gamma-p_2\cdot k\frac{\partial\Gamma}
{\partial p_1\cdot p_2}\right)+\frac{p_{2\mu}p_{2\nu}}{-p_2\cdot k}\left(
\Gamma-p_1\cdot k\frac{\partial\Gamma}{\partial p_1\cdot p_2}\right)\right.
\notag\\
&\left.\frac{p_{1(\mu}k_{\nu)}}{2p_1\cdot k}\Gamma+
\frac{p_{2(\mu}k_{\nu)}}{2p_2\cdot k}\Gamma\right]+\frac{1}{2}
\bar{\Gamma}^{\rm int.}_{\mu\nu},
\label{gam3}
\end{align}
where $\Gamma\equiv\Gamma(p_1\cdot p_2)$, and we have used the notation
$p_{i(\mu}k_{\nu)}=p_{i\mu}k_\nu+p_{i\nu}k_\mu$. One may now use the 
gravitational Ward identity~\cite{Englert}
\begin{equation}
k^\mu\bar{\Gamma}_{\mu\nu}=0
\label{ward}
\end{equation}
(which follows from general coordinate invariance) as well as the on-shell
condition for the emitted graviton ($k^2=0$) to find
\begin{align}
\frac{1}{2}k^\mu
\bar{\Gamma}^{\rm int.}_{\mu\nu}+\frac{\kappa}{2}\left[-p_{1\nu}\left(
\Gamma-p_2\cdot k\frac{\partial\Gamma}{\partial p_1\cdot p_2}\right)
-p_{2\nu}\left(\Gamma-p_1\cdot k\frac{\partial\Gamma}{\partial p_1\cdot p_2}
\right)\right]=0.
\label{gam4}
\end{align}
The zeroth order term vanishes from momentum conservation ($p_1+p_2=0$).
Also, one may identify
\begin{equation}
p_{2\mu}\frac{\partial\Gamma}{\partial p_1\cdot p_2}=\frac{\partial\Gamma}
{\partial p_1^\mu},\quad p_{1\mu}\frac{\partial\Gamma}{\partial p_1\cdot p_2}
=\frac{\partial\Gamma}{\partial p_2^\mu},
\label{derivs}
\end{equation}
so that eq.~(\ref{gam4}) becomes
\begin{equation}
\bar{\Gamma}^{\rm int.}_{\mu\nu}=-\kappa\left[p_{1\nu}\frac{\partial\Gamma}
{\partial p_1^\mu}+p_{2\nu}\frac{\partial\Gamma}{\partial p_2^\mu}\right].
\label{gamresult}
\end{equation}
This is indeed a special case of eq.~(\ref{gravlow}) for internal emission 
graphs.

\bibliographystyle{JHEP}
\bibliography{refs.bib}

\providecommand{\href}[2]{#2}\begingroup\raggedright\begin{thebibliography}{10}

\bibitem{Hamber:2007fk}
H.~W. Hamber, {\it {Discrete and continuum quantum gravity}},
  \href{http://xxx.lanl.gov/abs/0704.2895}{{\tt 0704.2895}}.

\bibitem{Hohm:2011dz}
O.~Hohm, {\it {On factorizations in perturbative quantum gravity}},
  \href{http://xxx.lanl.gov/abs/1103.0032}{{\tt 1103.0032}}.

\bibitem{Bern:2010yg}
Z.~Bern, T.~Dennen, Y.-t. Huang, and M.~Kiermaier, {\it {Gravity as the Square
  of Gauge Theory}},  {\em Phys.Rev.} {\bf D82} (2010) 065003,
  [\href{http://xxx.lanl.gov/abs/1004.0693}{{\tt 1004.0693}}].

\bibitem{Bern:2010ue}
Z.~Bern, J.~J.~M. Carrasco, and H.~Johansson, {\it {Perturbative Quantum
  Gravity as a Double Copy of Gauge Theory}},  {\em Phys.Rev.Lett.} {\bf 105}
  (2010) 061602, [\href{http://xxx.lanl.gov/abs/1004.0476}{{\tt 1004.0476}}].

\bibitem{Bern:2002kj}
Z.~Bern, {\it {Perturbative quantum gravity and its relation to gauge theory}},
   {\em Living Rev.Rel.} {\bf 5} (2002) 5,
  [\href{http://xxx.lanl.gov/abs/gr-qc/0206071}{{\tt gr-qc/0206071}}].

\bibitem{Bern:1999ji}
Z.~Bern and A.~K. Grant, {\it {Perturbative gravity from QCD amplitudes}},
  {\em Phys.Lett.} {\bf B457} (1999) 23--32,
  [\href{http://xxx.lanl.gov/abs/hep-th/9904026}{{\tt hep-th/9904026}}].

\bibitem{ArkaniHamed:2008yf}
N.~Arkani-Hamed and J.~Kaplan, {\it {On Tree Amplitudes in Gauge Theory and
  Gravity}},  {\em JHEP} {\bf 0804} (2008) 076,
  [\href{http://xxx.lanl.gov/abs/0801.2385}{{\tt 0801.2385}}].

\bibitem{Bern:1998ug}
Z.~Bern, L.~J. Dixon, D.~Dunbar, M.~Perelstein, and J.~Rozowsky, {\it {On the
  relationship between Yang-Mills theory and gravity and its implication for
  ultraviolet divergences}},  {\em Nucl.Phys.} {\bf B530} (1998) 401--456,
  [\href{http://xxx.lanl.gov/abs/hep-th/9802162}{{\tt hep-th/9802162}}].

\bibitem{Bern:2006kd}
Z.~Bern, L.~J. Dixon, and R.~Roiban, {\it {Is N = 8 supergravity ultraviolet
  finite?}},  {\em Phys.Lett.} {\bf B644} (2007) 265--271,
  [\href{http://xxx.lanl.gov/abs/hep-th/0611086}{{\tt hep-th/0611086}}].

\bibitem{Bern:2007hh}
Z.~Bern, J.~Carrasco, L.~J. Dixon, H.~Johansson, D.~Kosower, {\em et~al.}, {\it
  {Three-Loop Superfiniteness of N=8 Supergravity}},  {\em Phys.Rev.Lett.} {\bf
  98} (2007) 161303, [\href{http://xxx.lanl.gov/abs/hep-th/0702112}{{\tt
  hep-th/0702112}}].

\bibitem{Bern:2008pv}
Z.~Bern, J.~Carrasco, L.~J. Dixon, H.~Johansson, and R.~Roiban, {\it {Manifest
  Ultraviolet Behavior for the Three-Loop Four-Point Amplitude of N=8
  Supergravity}},  {\em Phys.Rev.} {\bf D78} (2008) 105019,
  [\href{http://xxx.lanl.gov/abs/0808.4112}{{\tt 0808.4112}}].

\bibitem{Bern:2009kd}
Z.~Bern, J.~Carrasco, L.~J. Dixon, H.~Johansson, and R.~Roiban, {\it {The
  Ultraviolet Behavior of N=8 Supergravity at Four Loops}},  {\em
  Phys.Rev.Lett.} {\bf 103} (2009) 081301,
  [\href{http://xxx.lanl.gov/abs/0905.2326}{{\tt 0905.2326}}].

\bibitem{Naculich:2011ry}
S.~G. Naculich and H.~J. Schnitzer, {\it {Absence of subleading infrared
  divergences in gravity}},  \href{http://xxx.lanl.gov/abs/1101.1524}{{\tt
  1101.1524}}. * Temporary entry *.

\bibitem{Bloch:1937pw}
F.~Bloch and A.~Nordsieck, {\it {Note on the Radiation Field of the electron}},
   {\em Phys.Rev.} {\bf 52} (1937) 54--59.

\bibitem{Yennie:1961ad}
D.~Yennie, S.~C. Frautschi, and H.~Suura, {\it {The infrared divergence
  phenomena and high-energy processes}},  {\em Annals Phys.} {\bf 13} (1961)
  379--452.

\bibitem{Sterman:1986aj}
G.~F. Sterman, {\it {Summation of Large Corrections to Short Distance Hadronic
  Cross-Sections}},  {\em Nucl.Phys.} {\bf B281} (1987) 310.

\bibitem{Catani:1989ne}
S.~Catani and L.~Trentadue, {\it {Resummation of the QCD Perturbative Series
  for Hard Processes}},  {\em Nucl.Phys.} {\bf B327} (1989) 323.

\bibitem{Korchemsky:1992xv}
G.~Korchemsky and G.~Marchesini, {\it {Structure function for large x and
  renormalization of Wilson loop}},  {\em Nucl.Phys.} {\bf B406} (1993)
  225--258, [\href{http://xxx.lanl.gov/abs/hep-ph/9210281}{{\tt
  hep-ph/9210281}}].

\bibitem{Korchemsky:1993uz}
G.~Korchemsky and G.~Marchesini, {\it {Resummation of large infrared
  corrections using Wilson loops}},  {\em Phys.Lett.} {\bf B313} (1993)
  433--440.

\bibitem{Beneke:2002ph}
M.~Beneke, A.~Chapovsky, M.~Diehl, and T.~Feldmann, {\it {Soft collinear
  effective theory and heavy to light currents beyond leading power}},  {\em
  Nucl.Phys.} {\bf B643} (2002) 431--476,
  [\href{http://xxx.lanl.gov/abs/hep-ph/0206152}{{\tt hep-ph/0206152}}].

\bibitem{Bauer:2000yr}
C.~W. Bauer, S.~Fleming, D.~Pirjol, and I.~W. Stewart, {\it {An Effective field
  theory for collinear and soft gluons: Heavy to light decays}},  {\em
  Phys.Rev.} {\bf D63} (2001) 114020,
  [\href{http://xxx.lanl.gov/abs/hep-ph/0011336}{{\tt hep-ph/0011336}}].

\bibitem{Bauer:2002nz}
C.~W. Bauer, S.~Fleming, D.~Pirjol, I.~Z. Rothstein, and I.~W. Stewart, {\it
  {Hard scattering factorization from effective field theory}},  {\em
  Phys.Rev.} {\bf D66} (2002) 014017,
  [\href{http://xxx.lanl.gov/abs/hep-ph/0202088}{{\tt hep-ph/0202088}}].

\bibitem{Becher:2006nr}
T.~Becher and M.~Neubert, {\it {Threshold resummation in momentum space from
  effective field theory}},  {\em Phys.Rev.Lett.} {\bf 97} (2006) 082001,
  [\href{http://xxx.lanl.gov/abs/hep-ph/0605050}{{\tt hep-ph/0605050}}].

\bibitem{Becher:2006mr}
T.~Becher, M.~Neubert, and B.~D. Pecjak, {\it {Factorization and Momentum-Space
  Resummation in Deep-Inelastic Scattering}},  {\em JHEP} {\bf 0701} (2007)
  076, [\href{http://xxx.lanl.gov/abs/hep-ph/0607228}{{\tt hep-ph/0607228}}].

\bibitem{Becher:2007ty}
T.~Becher, M.~Neubert, and G.~Xu, {\it {Dynamical Threshold Enhancement and
  Resummation in Drell-Yan Production}},  {\em JHEP} {\bf 0807} (2008) 030,
  [\href{http://xxx.lanl.gov/abs/0710.0680}{{\tt 0710.0680}}].

\bibitem{Laenen:2008gt}
E.~Laenen, G.~Stavenga, and C.~D. White, {\it {Path integral approach to
  eikonal and next-to-eikonal exponentiation}},  {\em JHEP} {\bf 0903} (2009)
  054, [\href{http://xxx.lanl.gov/abs/0811.2067}{{\tt 0811.2067}}].

\bibitem{Mueller:1979ih}
A.~H. Mueller, {\it On the asymptotic behavior of the {S}udakov form-factor},
  {\em Phys.Rev.} {\bf D20} (1979) 2037.

\bibitem{Collins:1980ih}
J.~C. Collins, {\it Algorithm to compute corrections to the {S}udakov
  form-factor},  {\em Phys.Rev.} {\bf D22} (1980) 1478.

\bibitem{Sen:1981sd}
A.~Sen, {\it {Asymptotic Behavior of the Sudakov Form-Factor in QCD}},  {\em
  Phys.Rev.} {\bf D24} (1981) 3281.

\bibitem{Korchemsky:1988pn}
G.~Korchemsky, {\it Double logarithmic asymptotics in {QCD}},  {\em Phys.Lett.}
  {\bf B217} (1989) 330--334.

\bibitem{Magnea:1990zb}
L.~Magnea and G.~F. Sterman, {\it {Analytic continuation of the Sudakov
  form-factor in QCD}},  {\em Phys.Rev.} {\bf D42} (1990) 4222--4227.

\bibitem{Weinberg:1965nx}
S.~Weinberg, {\it {Infrared photons and gravitons}},  {\em Phys.Rev.} {\bf 140}
  (1965) B516--B524.

\bibitem{Dixon:2008gr}
L.~J. Dixon, L.~Magnea, and G.~F. Sterman, {\it {Universal structure of
  subleading infrared poles in gauge theory amplitudes}},  {\em JHEP} {\bf
  0808} (2008) 022, [\href{http://xxx.lanl.gov/abs/0805.3515}{{\tt
  0805.3515}}].

\bibitem{Gardi:2009qi}
E.~Gardi and L.~Magnea, {\it {Factorization constraints for soft anomalous
  dimensions in QCD scattering amplitudes}},  {\em JHEP} {\bf 0903} (2009) 079,
  [\href{http://xxx.lanl.gov/abs/0901.1091}{{\tt 0901.1091}}].

\bibitem{Dixon:2009gx}
L.~J. Dixon, {\it {Matter Dependence of the Three-Loop Soft Anomalous Dimension
  Matrix}},  {\em Phys.Rev.} {\bf D79} (2009) 091501,
  [\href{http://xxx.lanl.gov/abs/0901.3414}{{\tt 0901.3414}}].

\bibitem{Becher:2009cu}
T.~Becher and M.~Neubert, {\it {Infrared singularities of scattering amplitudes
  in perturbative QCD}},  {\em Phys.Rev.Lett.} {\bf 102} (2009) 162001,
  [\href{http://xxx.lanl.gov/abs/0901.0722}{{\tt 0901.0722}}].

\bibitem{Becher:2009qa}
T.~Becher and M.~Neubert, {\it {On the Structure of Infrared Singularities of
  Gauge-Theory Amplitudes}},  {\em JHEP} {\bf 0906} (2009) 081,
  [\href{http://xxx.lanl.gov/abs/0903.1126}{{\tt 0903.1126}}].

\bibitem{Becher:2009kw}
T.~Becher and M.~Neubert, {\it {Infrared singularities of QCD amplitudes with
  massive partons}},  {\em Phys.Rev.} {\bf D79} (2009) 125004,
  [\href{http://xxx.lanl.gov/abs/0904.1021}{{\tt 0904.1021}}].

\bibitem{Dixon:2009ur}
L.~J. Dixon, E.~Gardi, and L.~Magnea, {\it {On soft singularities at three
  loops and beyond}},  {\em JHEP} {\bf 1002} (2010) 081,
  [\href{http://xxx.lanl.gov/abs/0910.3653}{{\tt 0910.3653}}].

\bibitem{Dixon:2010zz}
L.~J. Dixon, E.~Gardi, and L.~Magnea, {\it {All-order results for infrared and
  collinear singularities in massless gauge theories}},  {\em PoS} {\bf
  RADCOR2009} (2010) 007, [\href{http://xxx.lanl.gov/abs/1001.4709}{{\tt
  1001.4709}}]. * Temporary entry *.

\bibitem{Gardi:2009zv}
E.~Gardi and L.~Magnea, {\it {Infrared singularities in QCD amplitudes}},  {\em
  Nuovo Cim.} {\bf C32N5-6} (2009) 137--157,
  [\href{http://xxx.lanl.gov/abs/0908.3273}{{\tt 0908.3273}}].

\bibitem{Laenen:2010uz}
E.~Laenen, L.~Magnea, G.~Stavenga, and C.~D. White, {\it {Next-to-eikonal
  corrections to soft gluon radiation: a diagrammatic approach}},
  \href{http://xxx.lanl.gov/abs/1010.1860}{{\tt 1010.1860}}.

\bibitem{Gatheral:1983cz}
J.~Gatheral, {\it Exponentiation of eikonal cross-sections in nonabelian gauge
  theories},  {\em Phys.Lett.} {\bf B133} (1983) 90.

\bibitem{Frenkel:1984pz}
J.~Frenkel and J.~C. Taylor, {\it Nonabelian eikonal exponentiation},  {\em
  Nucl. Phys.} {\bf B246} (1984) 231.

\bibitem{Sterman:1981jc}
G.~Sterman, {\it Infrared divergences in perturbative {QCD}}, . in
  {T}allahassee 1981, Proceedings, Perturbative Quantum Chromodynamics, 22-40.

\bibitem{Gardi:2010rn}
E.~Gardi, E.~Laenen, G.~Stavenga, and C.~D. White, {\it {Webs in multiparton
  scattering using the replica trick}},  {\em JHEP} {\bf 1011} (2010) 155,
  [\href{http://xxx.lanl.gov/abs/1008.0098}{{\tt 1008.0098}}].

\bibitem{Gardi:2011wa}
E.~Gardi and C.~D. White, {\it {General properties of multiparton webs: proofs
  from combinatorics}},  \href{http://xxx.lanl.gov/abs/1102.0756}{{\tt
  1102.0756}}. * Temporary entry *.

\bibitem{Mitov:2010rp}
A.~Mitov, G.~Sterman, and I.~Sung, {\it {Diagrammatic Exponentiation for
  Products of Wilson Lines}},  {\em Phys.Rev.} {\bf D82} (2010) 096010,
  [\href{http://xxx.lanl.gov/abs/1008.0099}{{\tt 1008.0099}}].

\bibitem{Laenen:2008ux}
E.~Laenen, L.~Magnea, and G.~Stavenga, {\it {On next-to-eikonal corrections to
  threshold resummation for the Drell-Yan and DIS cross sections}},  {\em
  Phys.Lett.} {\bf B669} (2008) 173--179,
  [\href{http://xxx.lanl.gov/abs/0807.4412}{{\tt 0807.4412}}].

\bibitem{Moch:2009hr}
S.~Moch and A.~Vogt, {\it {On non-singlet physical evolution kernels and
  large-x coefficient functions in perturbative QCD}},  {\em JHEP} {\bf 11}
  (2009) 099, [\href{http://xxx.lanl.gov/abs/0909.2124}{{\tt 0909.2124}}].

\bibitem{Soar:2009yh}
G.~Soar, S.~Moch, J.~A.~M. Vermaseren, and A.~Vogt, {\it {On Higgs-exchange
  DIS, physical evolution kernels and fourth-order splitting functions at large
  x}},  {\em Nucl. Phys.} {\bf B832} (2010) 152,
  [\href{http://xxx.lanl.gov/abs/0912.0369}{{\tt 0912.0369}}].

\bibitem{Vogt:2010pe}
A.~Vogt, G.~Soar, S.~Moch, and J.~Vermaseren, {\it {On higher-order
  flavour-singlet splitting and coefficient functions at large x}},
  \href{http://xxx.lanl.gov/abs/1008.0952}{{\tt 1008.0952}}.

\bibitem{Vogt:2010ik}
A.~Vogt, S.~Moch, G.~Soar, and J.~A.~M. Vermaseren, {\it {Higher-order
  predictions from physical evolution kernels}},  {\em PoS} {\bf RADCOR2009}
  (2010) 053, [\href{http://xxx.lanl.gov/abs/1001.3554}{{\tt 1001.3554}}].

\bibitem{Grunberg:2009yi}
G.~Grunberg and V.~Ravindran, {\it {On threshold resummation beyond leading
  $1-x$ order}},  {\em JHEP} {\bf 10} (2009) 055,
  [\href{http://xxx.lanl.gov/abs/0902.2702}{{\tt 0902.2702}}].

\bibitem{Grunberg:2009vs}
G.~Grunberg, {\it {Large-$x$ structure of physical evolution kernels in Deep
  Inelastic Scatterin\ g}},  \href{http://xxx.lanl.gov/abs/0911.4471}{{\tt
  0911.4471}}.

\bibitem{Grunberg:2010sw}
G.~Grunberg, {\it {Threshold resummation beyond leading eikonal level}},
  \href{http://xxx.lanl.gov/abs/1005.5684}{{\tt 1005.5684}}.

\bibitem{Low:1958sn}
F.~E. Low, {\it {Bremsstrahlung of very low-energy quanta in elementary
  particle collisions}},  {\em Phys. Rev.} {\bf 110} (1958) 974--977.

\bibitem{Burnett:1967km}
T.~H. Burnett and N.~M. Kroll, {\it {Extension of the low soft photon
  theorem}},  {\em Phys. Rev. Lett.} {\bf 20} (1968) 86.

\bibitem{DelDuca:1990gz}
V.~Del~Duca, {\it High-energy bremsstrahlung theorems for soft photons},  {\em
  Nucl. Phys.} {\bf B345} (1990) 369--388.

\bibitem{Giddings:2010pp}
S.~B. Giddings, M.~Schmidt-Sommerfeld, and J.~R. Andersen, {\it {High energy
  scattering in gravity and supergravity}},  {\em Phys.Rev.} {\bf D82} (2010)
  104022, [\href{http://xxx.lanl.gov/abs/1005.5408}{{\tt 1005.5408}}].

\bibitem{Stirling:2011mf}
W.~Stirling, E.~Vryonidou, and J.~Wells, {\it {Eikonal regime of
  gravity-induced scattering at higher energy proton colliders}},
  \href{http://xxx.lanl.gov/abs/1102.3844}{{\tt 1102.3844}}. * Temporary entry
  *.

\bibitem{Capper:1973bk}
D.~Capper, {\it On quantum corrections to the graviton propagator},  {\em Nuovo
  Cim.} {\bf A25} (1975) 29.

\bibitem{Capper:1973pv}
D.~Capper, G.~Leibbrandt, and M.~Ramon~Medrano, {\it {Calculation of the
  graviton selfenergy using dimensional regularization}},  {\em Phys.Rev.} {\bf
  D8} (1973) 4320--4331.

\bibitem{Strassler:1992zr}
M.~J. Strassler, {\it {Field theory without Feynman diagrams: One loop
  effective actions}},  {\em Nucl.Phys.} {\bf B385} (1992) 145--184,
  [\href{http://xxx.lanl.gov/abs/hep-ph/9205205}{{\tt hep-ph/9205205}}].

\bibitem{Schmidt:1994zj}
M.~G. Schmidt and C.~Schubert, {\it {Worldline Green functions for multiloop
  diagrams}},  {\em Phys.Lett.} {\bf B331} (1994) 69--76,
  [\href{http://xxx.lanl.gov/abs/hep-th/9403158}{{\tt hep-th/9403158}}].

\bibitem{vanHolten:1995ds}
J.~van Holten, {\it {Propagators and path integrals}},  {\em Nucl.Phys.} {\bf
  B457} (1995) 375--407, [\href{http://xxx.lanl.gov/abs/hep-th/9508136}{{\tt
  hep-th/9508136}}].

\bibitem{Bern:1991an}
Z.~Bern and D.~C. Dunbar, {\it {A Mapping between Feynman and string motivated
  one loop rules in gauge theories}},  {\em Nucl.Phys.} {\bf B379} (1992)
  562--601.

\bibitem{Bern:1991aq}
Z.~Bern and D.~A. Kosower, {\it {The Computation of loop amplitudes in gauge
  theories}},  {\em Nucl.Phys.} {\bf B379} (1992) 451--561.

\bibitem{Karanikas:2002sy}
A.~Karanikas, C.~Ktorides, and N.~Stefanis, {\it {World line techniques for
  resumming gluon radiative corrections at the cross-section level}},  {\em
  Eur.Phys.J.} {\bf C26} (2003) 445--455,
  [\href{http://xxx.lanl.gov/abs/hep-ph/0210042}{{\tt hep-ph/0210042}}].

\bibitem{Sterman:1995fz}
G.~F. Sterman, {\it {Partons, factorization and resummation, TASI 95}},
  \href{http://xxx.lanl.gov/abs/hep-ph/9606312}{{\tt hep-ph/9606312}}.

\bibitem{Morgan:1995te}
A.~Morgan, {\it {Second order fermions in gauge theories}},  {\em Phys.Lett.}
  {\bf B351} (1995) 249--256,
  [\href{http://xxx.lanl.gov/abs/hep-ph/9502230}{{\tt hep-ph/9502230}}].

\bibitem{Binosi:2008ig}
D.~Binosi, J.~Collins, C.~Kaufhold, and L.~Theussl, {\it {JaxoDraw: A graphical
  user interface for drawing Feynman diagrams. Version 2.0 release notes}},
  {\em Comput. Phys. Commun.} {\bf 180} (2009) 1709--1715,
  [\href{http://xxx.lanl.gov/abs/0811.4113}{{\tt 0811.4113}}].

\bibitem{Binosi:2003yf}
D.~Binosi and L.~Theussl, {\it {JaxoDraw: A graphical user interface for
  drawing Feynman diagrams}},  {\em Comput. Phys. Commun.} {\bf 161} (2004)
  76--86, [\href{http://xxx.lanl.gov/abs/hep-ph/0309015}{{\tt
  hep-ph/0309015}}].

\bibitem{Englert}
F.~Englert and R.~Brout, {\it {Gravitational Ward Identity and the Principle of
  Equivalence}},  {\em Phys. Rev.} {\bf 141} (1965) 1231.

\end{thebibliography}\endgroup
\end{document}